\documentclass[12pt,a4paper]{article}
\usepackage{jheppub}
\usepackage{caption}

\usepackage{graphicx, epsfig}
\usepackage{color}
\usepackage{mathrsfs}
\usepackage{amsmath}
\usepackage{bbold}
\usepackage{amssymb}

\newcommand{\Vol}{\mathcal{V}}
\newcommand{\gravm}{m_{\text{cs},\,3/2}}

\newcommand{\Pex}{\mathbb{CP}^4_{11169}[18]}
\newcommand{\be}{\begin{equation}}
\newcommand{\ee}{\end{equation}}
\newcommand{\bea}{\begin{eqnarray}}
\newcommand{\eea}{\end{eqnarray}}
\newcommand{\barr}{\begin{array}}
\newcommand{\earr}{\end{array}}

\def\beq{\begin{equation}}
\def\eeq{\end{equation}}
\def\be{\begin{equation}}
\def\ee{\end{equation}}
\def\bea{\begin{eqnarray}}
\def\eea{\end{eqnarray}}

\title{Finding all flux vacua in an explicit example}
\author[a]{Danny~Mart\'inez-Pedrera,} \author[b]{Dhagash~Mehta,} \author[a]{Markus~Rummel,} \author[c]{and Alexander~Westphal}

\affiliation[a]{II. Institut f\"ur Theoretische Physik der Universit\"at Hamburg, D-22761 Hamburg, Germany}
\affiliation[b]{Department of Physics, Syracuse University, Syracuse, NY 13244, USA}
\affiliation[c]{Deutsches Elektronen-Synchrotron DESY, Theory Group, D-22603 Hamburg, Germany}
\emailAdd{dbmehta@syr.edu}\emailAdd{markus.rummel@desy.de}\emailAdd{alexander.westphal@desy.de}
\abstract{We explicitly construct all supersymmetric flux vacua of a particular Calabi-Yau compactification of type IIB string theory for a small number of flux carrying cycles and a given D3-brane tadpole. The analysis is performed in the large complex structure region by using the polynomial homotopy continuation method, which allows to find all stationary points of the polynomial equations that characterize the supersymmetric vacuum solutions. The number of vacua as a function of the D3 tadpole is in agreement with statistical studies in the literature. We calculate the available tuning of the cosmological constant from fluxes and extrapolate to scenarios with a larger number of flux carrying cycles. We also verify the range of scales for the moduli and gravitino masses recently found for a single explicit flux choice giving a K\"ahler uplifted de Sitter vacuum in the same construction.}

\begin{document}
\begin{flushright}
DESY 12-246\\[0.7cm]
\end{flushright}
\maketitle
\setlength{\parskip}{0.2cm}

\section{Introduction}

The question of the vacuum states of string theory has been an outstanding problem in the field since its founding days. Those vacua which could describe our world are of particular interest. Apart from a standard model sector including the corresponding gauge groups and matter in the right representations of these gauge groups, such vacua should have a positive vacuum energy that is extremely small in terms of the natural scale of gravity, the Planck / string scale~\cite{Riess:1998cb,Perlmutter:1998np,Komatsu:2010fb}. In addition, one may wish for 4D $\mathcal{N}=1$ supersymmetry for its power to keep control over quantum corrections to the theory above the supersymmetry breaking / electroweak scale. There has been remarkable progress in the construction of such de Sitter (dS) vacua with stabilized geometric moduli in the past years~\cite{Dasgupta:1999ss,Giddings:2001yu,Kachru:2003aw}. The moduli are 4D massless scalar fields parametrizing the geometric deformation modes of the compact six-dimensional spaces all viable Kaluza-Klein type vacua of string theory need to describe our effectively four-dimensional reality.

A corner of the string theory landscape where moduli stabilization can be addressed very explicitly is type IIB compactified on orientifolded Calabi-Yau threefolds. The four-dimensional effective action of the geometric moduli is given by an $\mathcal{N}=1$ theory of a set of chiral multiplets consisting of the axio-dilaton, $h^{1,1}$ K\"ahler moduli, and $h^{2,1}$ complex structure moduli~\cite{Grimm:2004uq}. Recent years saw a progress for the stabilization of the dilaton and the complex structure moduli from the use of  quantized fluxes of three form field strength of the Ramond-Ramond and Neveu-Schwarz sector of type IIB string theory~\cite{Giddings:2001yu}. The flux stabilization procedure operates supersymmetrically at a high scale. The K\"ahler moduli are flat directions at tree level, i.e. they do not obtain a scalar potential due to the no-scale structure of type IIB compactified on Calabi-Yau~\cite{Ellis:1983sf,Cremmer:1983bf}. We can use the breaking of this no-scale structure by non-perturbative~\cite{Kachru:2003aw} and perturbative~\cite{Becker:2002nn,Berg:2005ja} effects to stabilize the K\"ahler moduli at a parametrically lower scale than the complex structure moduli. This stabilization produces in an AdS vacuum with unbroken supersymmetry~\cite{Kachru:2003aw}, an AdS vacuum with spontaneously broken supersymmetry with an exponentially large volume~\cite{Balasubramanian:2004uy,Balasubramanian:2005zx} or directly in a dS vacuum with spontaneously broken supersymmetry~\cite{Rummel:2011cd,Louis:2012nb}.

 If the stabilization of the geometric moduli does not directly lead to a dS vacuum, an additional uplifting sector has to be added. The uplifting may arise for instance by anti D3 branes~\cite{Kachru:2003aw}, D-terms~\cite{Burgess:2003ic,Haack:2006cy,Cicoli:2012vw}, F-terms from matter fields~\cite{Lebedev:2006qq}, metastable vacua in gauge theories~\cite{Intriligator:2006dd} or dilaton dependent non-perturbative effects~\cite{Cicoli:2012fh}. Concerning the smallness of the cosmological constant,  statistical arguments show that due to the enormous amount of possible flux configurations there is an exponential abundance of isolated potential dS vacua~\cite{Bousso:2000xa,Feng:2000if,Denef:2004ze}. For a sufficiently large number of complex structure moduli $h^{2,1}$ (typically ${\cal O}(100)$) the number of flux vacua and, in turn, also the number of dS vacua scaling like $e^{{\cal O}(1)\, h^{2,1}}$ is large enough to produce vacuum energies with average spacing $\lesssim 10^{-120}$. This enables the flux vacuum landscape in string theory in principle to accommodate the observed vacuum energy of our Universe.

In this paper, we study the flux vacua of a particular Calabi-Yau: The degree 18 hypersurface in a 4 complex dimensional projective space $X_3 \equiv \Pex$ in the large complex structure limit. This manifold is the standard working example of both the large volume scenario (LVS)~\cite{Balasubramanian:2004uy,Balasubramanian:2005zx} and the K\"ahler uplifting scenario~\cite{Rummel:2011cd,Louis:2012nb} and its geometric properties have been worked out in great detail in~\cite{Candelas:1994hw}. It has $h^{1,1}=2$ K\"ahler moduli and $h^{2,1}=272$ complex structure moduli. We switch on flux along six three-cycles that correspond to two complex structure moduli that are invariant under a certain discrete symmetry that can be used to construct the mirror manifold~\cite{Greene:1990ud}. For this purpose we review a known argument that a supersymmetric vacuum in these two complex structure moduli corresponds to a supersymmetric vacuum of \textit{all} 272 complex structure moduli~\cite{DeWolfe:2005uu,Denef:2004dm}.

For an explicit construction of the flux vacua we use the fact that the prepotential $\mathcal{G}$ of the two complex structure moduli space has been worked out in~\cite{Candelas:1994hw} in the large complex structure limit. We apply two computational methods to find flux vacua on this manifold:
\begin{itemize}
 \item \textit{The polynomial homotopy continuation method}~\cite{SW:95} allows us to find \textit{all} stationary points of the polynomial equations that characterize the supersymmetric vacuum solutions. The fluxes $f_i \in \mathbb{Z}$ appear as parameters in these equations and are restricted by the D3 tadpole $L$ which depends on the chosen brane and gauge flux configuration imposed on the manifold. Since the restriction is of the form $\sum f_i^2 \leq L$ this method allows us to explicitly construct for the first time \textit{all flux vacua} in the large complex structure limit that are consistent with a given D3 tadpole $L$ by applying the polynomial homotopy continuation method at each point in flux parameter space. This method has the attractive feature to be highly parallelizable.
 \item \textit{The minimal flux method}~\cite{Denef:2004dm} finds flux parameters that are consistent with a given D3 tadpole $L$ for a given set of vacuum expectation values (VEVs) of the complex structure moduli. Hence it is in a sense complementary to the polynomial homotopy continuation method where the role of parameters and solutions is exchanged with respect to this method. However, it is not possible to find all flux vacua for a given tadpole $L$ with this method.
\end{itemize}

The obtained solution space of flux vacua is analyzed for several physically interesting properties. For the polynomial homotopy continuation method, we find that for the $\sim 50,000$ flux choices contained in our maximum D3-brane tadpole $L=34$ of our scan there are $\sim 20,000$ solutions in the large complex structure limit. We find a preference of strongly coupled vacua $g_s \gtrsim 1$ and preference for values of $\mathcal{O}(10^1-10^3)$ for the flux superpotential $W_0$. The number of vacua is
\begin{equation}
N_{vac} \simeq (0.52\pm 0.04)\, L^{2.92\pm 0.03} \,,
\end{equation}
compared to $\sim 0.03 L^3$ expected form statistical analysis~\cite{Ashok:2003gk,Denef:2004ze}. The gravitino mass is typically $m_{3/2}^2=\mathcal{O}(10^{-3}) \cdot(\frac{100}{\Vol})^{2} M_{\text{P}}^2$ and the masses of the complex structure moduli and the dilaton scale like $\mathcal{O}(10^{-3}-10^2) \cdot(\frac{100}{\Vol})^{2} M_{\text{P}}^2$, where $\Vol$ is the Volume of $X_3$ in string units.\\
The average spacing of the flux superpotential in our solution set can be used to estimate the available fine-tuning $\Delta \Lambda / \Lambda$ of the cosmological constant $\Lambda$ as
\begin{equation}
 \frac{\Delta \Lambda}{\Lambda} \simeq (5.1 \pm 0.3)\,L^{- (0.93\pm 0.006)\, ( h^{2,1}_{\text{eff}}+1)}\,,
 \label{DeltaLambdafitintro}
\end{equation}
where $h^{2,1}_{\text{eff}}$ is the number of complex structure moduli with non-zero flux on the corresponding three-cycles. Eq.~\eqref{DeltaLambdafitintro} is obtained as a fit for $L\leq 34$ and $h^{2,1}_{\text{eff}}=2$. It can be used to estimate the available fine-tuning of the cosmological constant to for instance $\Delta \Lambda / \Lambda \sim 10^{-100}$ for $L=500$ and $h^{2,1}_{\text{eff}} = 40$. These are typical values for Calabi-Yau manifolds that are hypersurfaces in toric varieties with D7 branes and O7 planes introduced to stabilize the K\"ahler moduli and break supersymmetry.\\
The explicit brane and gauge flux construction in~\cite{Louis:2012nb} allows us to answer the question how many of theses supersymmetric flux vacua allow an uplift to dS via K\"ahler uplifting. Depending on the available values for the one-loop determinant from gaugino condensation used to stabilize the K\"ahler moduli in this setup, we find that for a fraction of about $10^{-4}$ of all flux vacua up to a given D3-brane tadpole this mechanism can be applied to obtain a dS vacuum.

For the minimal flux method, we find $\sim 1000$ flux vacua with $L<500$ out of $\sim 10^7$ parameter points of our scan. This method allows us to control the region in $W_0$ and moduli space where we are intending to find flux vacua. Hence, we more easily access the regions of weak string coupling and the large complex structure limit compared to the polynomial homotopy continuation method. For this much smaller set of flux vacua constructed with the minimal flux method, the fraction of K\"ahler uplifted dS minima is about $10\%$. This is a considerably higher fraction of vacua compared to the polynomial homotopy continuation method which is due to the fact that the minimal flux method naturally finds values for $W_0$ in a region where K\"ahler uplifting is applicable.

Our results are complementary to statistical analysis by~\cite{Ashok:2003gk,Denef:2004ze}.\footnote{For explicitly constructed vacua on two different two parameter models in the vicinity of the Landau-Ginzburg respectively conifold point see~\cite{Conlon:2004ds}. For a study of flux vacua of $X_3$ in the context of accidental inflation~\cite{Linde:2007jn} see \cite{BlancoPillado:2012cb}.} The uniform distribution of physical quantities as for instance the gravitino mass and the vacuum energy density in the landscape has recently been questioned in general~\cite{Marsh:2011aa,Chen:2011ac,Bachlechner:2012at} and in the context of K\"ahler uplifting~\cite{Sumitomo:2012wa,Sumitomo:2012vx,Sumitomo:2012cf}. Hence, our results present an important check of the general results found in~\cite{Ashok:2003gk,Denef:2004ze} on a very realistic examples $X_3$, especially since we are able to construct the complete solution space of flux vacua for a given tadpole $L$.

In section~\ref{complexstructure_sec}, we review the effective 4d, $\mathcal{N}=1$ description of the complex structure moduli and the dilaton as well as the reduction of the full moduli space to two complex structure moduli. The scans for flux vacua with the polynomial homotopy continuation method and the minimal flux method are presented in section~\ref{paramotopy_sec} respectively section~\ref{minflux_sec}. We conclude in section~\ref{conclusions}.

\section{The complex structure of $\Pex$} \label{complexstructure_sec}

The effective 4d, $\mathcal{N}=1$ description of the moduli space of the $h^{2,1} = 272$ complex structure moduli of $X_3$ is
given by the K\"ahler potential $K$ and superpotential $W_0$ of the theory.\footnote{For recent reviews see \cite{Douglas:2006es,Grana:2005jc,Blumenhagen:2006ci}.} We choose a symplectic basis $\{A_a,B^b\}$ for the $b_3 = 2 h^{2,1} + 2$ three-cycles
\begin{equation}
 \int_{A_a} \alpha^b = \int_{X_3} \alpha^b \wedge \beta_a = \delta_a^b\,,\qquad \qquad \int_{B^b} \beta_a = \int_{X_3} \beta_a \wedge \alpha^b = - \delta_a^b\,,
\end{equation}
where $\{\alpha_b,\beta^a\}$ are the Poincar\'e dual cohomology elements to the three-cycles and $a,b=0,..,h^{2,1}$.

Having chosen a symplectic basis for the three-cycles, this defines a choice of coordinates $\omega_a$ on complex structure moduli space via the period integrals over the holomorphic three-form $\Omega$ via
\begin{equation}
 \omega_a = \int_{A_a} \Omega\,,\qquad\qquad \mathcal{G}_b = \int_{B^b} \Omega\,.\label{defperiods}
\end{equation}
Note, that there are $h^{2,1}+1$ coordinates $\omega_0,..,\omega_{h^{2,1}}$ even though there are only $h^{2,1}$ complex structure moduli. This is because $\omega_0$ refers to the normalization of the holomorphic three-form $\Omega$. The complex structure moduli can be defined via $U_a \equiv \nu_a + i\, u_a = \omega_a / \omega_0$ for $i=a,..,h^{2,1}$. The period vector $\Pi(\omega_a)=(\mathcal{G}_b,\omega_a)$ is inherited from a holomorphic function $\mathcal{G}(\omega_a)$ of degree two in the $\omega_a$ known as the prepotential via $\mathcal{G}_b = \partial_b \mathcal{G}$ of the underlying ${\cal N}=2$ Calabi-Yau compactification.

The K\"ahler potential of the complex structure moduli $U_a$ and the dilaton $\tau= \sigma + i\, s$ can then be written as
\begin{align}
 \begin{aligned}\label{Kaehlercs}
  K_{\text{cs}} &= -\log \left(-i \int_{X_3} \Omega(U_a) \wedge \bar{\Omega}(\bar{U}_a) \right) - \log \left(-i(\tau-\bar \tau) \right)\,,\\
 &= -\log \left(i (\bar{\omega}_a \mathcal{G}_a - \omega_a \bar{\mathcal{G}}_a) \right) - \log \left(-i(\tau-\bar \tau) \right)\,,\\[2mm]
&= -\log \left(-i \Pi^\dagger \cdot \Sigma \cdot \Pi \right) - \log \left(-i(\tau-\bar \tau) \right)\,,
 \end{aligned}
\end{align}
where in the second line of eq.~\eqref{Kaehlercs} we have introduced the symplectic matrix
\begin{equation}
 \Sigma = \left( \begin{array}{cc}
           0 & \mathbb{1}\\
           -\mathbb{1} & 0
          \end{array} \right)\,,
\end{equation}
and used the intersection formula
\begin{equation}
 \int_{X_3} X \wedge Y = \sum_{a=0}^{h^{2,1}} \left( \int_{A_a} X \int_{B^a} Y - \int_{A_a} Y \int_{B^a} X \right)\,,\label{intformula}
\end{equation}
for general three-forms $X$ and $Y$.

The Gukov-Vafa-Witten flux superpotential is determined by the RR and NS flux $F_{3}$ and $H_{3}$ via~\cite{Gukov:1999ya}
\begin{equation}
 W_0 = \frac{1}{2\pi}\, \int_{X_3} (F_3 - \tau H_3) \wedge \Omega(U_a)\,.\label{W0gen}
\end{equation}
Due to the quantization of the three-form flux
\begin{align}
 \begin{aligned}
  &\frac{1}{(2\pi)^2 \alpha'} \int_{A_a} F_3 = {f_1}_a \in \mathbb{Z}\,,\qquad \frac{1}{(2\pi)^2 \alpha'} \int_{B^a} F_3 = {f_2}_a \in \mathbb{Z}\,,\\
 &\frac{1}{(2\pi)^2 \alpha'} \int_{A_a} H_3 = {h_1}_a \in \mathbb{Z}\,,\qquad \frac{1}{(2\pi)^2 \alpha'} \int_{B^a} H_3 = {h_2}_a \in \mathbb{Z}\,,
 \end{aligned}
\end{align}
eq.~\eqref{W0gen} can be written as
\begin{equation}
 W_0 = 2\pi \left[({f_1}_a-\tau\,{h_1}_a)\mathcal{G}_a- ({f_2}_a-\tau\,{h_2}_a)U_a\right]\,,\label{W0period}
\end{equation}
where we have set $\alpha'=1$ and used again eq.~\eqref{intformula} and the definition of the periods eq.~\eqref{defperiods}.

The D3-tadpole induced by turning on RR and NS flux is given by
\begin{equation}
 L = \frac{1}{(2\pi)^4 (\alpha')^2} \int_{X_3} H_3 \wedge F_3 =  h\cdot \Sigma \cdot f = h_1f_2-h_2f_1\,.\label{D3tad}
\end{equation}

The $\mathcal{N}=1$ supergravity scalar potential is given as
\begin{equation}
 V = e^{K} \left( K^{\alpha\bar{\beta}} D_\alpha W \overline{D_\beta W} - 3 |W|^2 \right)\,,\label{V4D}
\end{equation}
where $K=K_{\text{cs}}+K_{\text{k}}$ and
\begin{equation}
K_{\text{k}} = - 2 \log \Vol\,,\label{KaehlerK}
\end{equation}
is the K\"ahler potential of the K\"ahler moduli up to corrections in $\alpha'$ and $g_s$ with $\Vol$ the volume of the Calabi-Yau $X_3$. The indices $\alpha$ and $\beta$ in eq.~\eqref{V4D} run over the dilaton, the $h^{2,1}$ complex structure moduli and the $h^{1,1}$ K\"ahler moduli. At tree-level, eq.~\eqref{V4D} obeys a no-scale structure~\cite{Cremmer:1983bf,Ellis:1983sf} in the K\"ahler sector:
\begin{equation}
 K^{i \bar{\jmath}} D_i W \overline{D_j W} = 3 |W|^2\,, \qquad \text{for } i,j=1,..,h^{1,1}\,,\label{noscaleprop}
\end{equation}
such that
\begin{equation}
 V = e^{K} K^{c\bar{d}} D_c W \overline{D_d W}\,,\label{V4Dnoscale}
\end{equation}
where the indices $c$ and $d$ run over the moduli $\tau$ and $U_a$. The no-scale structure eq.~\eqref{noscaleprop} is broken by $\alpha'$ corrections~\cite{Becker:2002nn} and string loop corrections~\cite{Berg:2005ja} in $K$, as well as non-perturbative corrections to the superpotential and $K$. However, these corrections are parametrically small in every moduli stabilization scenario where the overall volume of $X_3$ is large. Hence, the scalar potential for the dilaton and complex structure moduli eq.~\eqref{V4Dnoscale} is positive semi-definite in this limit and a supersymmetric extremum given by a solution to the system of equations
\begin{equation}
 D_\tau W= 0 \qquad \text{and} \qquad D_{U_a} W = 0\,, \qquad \text{for } a=1,..,h^{2,1}\,,\label{DiWall}
\end{equation}
will always be a minimum, i.e. all eigenvalues of the second derivative matrix $V_{ab}$ are positive~\cite{Balasubramanian:2005zx}.

Note that due to the appearance of the symplectic matrix, the tadpole eq.~\eqref{D3tad} is at first not positive definite. However, as has been discussed in~\cite{Giddings:2001yu,Denef:2004ze}, imposing the supersymmetry conditions $D_a W = 0$ results in $G_3 = F_3 - \tau \,H_3$ being imaginary self-dual (ISD). Since the ISD component of $G_3$ always results in positive semi-definite contributions to the tadpole while the anti-ISD component of $G_3$ always yields negative semi-definite contributions, a supersymmetric point always has $L\geq 0$. This can be seen nicely if the ISD condition is displayed as~\cite{Ibanez:2012zz}
\begin{equation}
 \ast_6 s\,H_3 = -(F_3 - \sigma H_3)\,,\label{ISDF3H3}
\end{equation}
i.e. only $h^{2,1}+1$ of the original $2h^{2,1}+2$ fluxes are independent once the ISD condition is invoked and
\begin{equation}
 L \sim \int_{X_3} H_3 \wedge F_3 \sim \int_{X_3} H_3 \wedge \ast_6 H_3 \sim \int_{X_3} \sqrt{\tilde{g}} |H_3|^2>0\,,
\end{equation}
where we have used eq.~\eqref{ISDF3H3}.

The type IIB ten dimensional effective supergravity Lagrangian is invariant under $SL(2,\mathbb{Z})$ transformations
\begin{align}
 \begin{aligned}
  &\tau \rightarrow \frac{a \tau+b}{c \tau+d}\qquad \text{with} \qquad a,b,c,d\in \mathbb{Z}\quad\text{and}\quad ad-bc=1\,,\\[3mm]
&\left( \begin{array}{c}
         H_3\\ F_3
        \end{array}
 \right) \rightarrow
\left( \begin{array}{cc}
         d \,\,& c\\
         b \,\,& a
        \end{array}
 \right)
\cdot  \left( \begin{array}{c}
         H_3\\ F_3
        \end{array}
 \right)\,,\label{SL2Zgen}
 \end{aligned}
\end{align}
which implies
\begin{equation}
 G_3 \rightarrow \frac{G_3}{c \tau +d}\,.
\end{equation}
As is easily verified, these transformations also leave the D3 tadpole eq.~\eqref{D3tad} invariant. When determining the solution space of flux vacua of $X_3$ we have to make sure to consider only inequivalent vacua under the transformations eq.~\eqref{SL2Zgen}.

\subsection{Effective reduction of the moduli space}

Consider the two parameter $\psi, \phi$-family of threefolds $\Pex$ given by the vanishing of the polynomials
\begin{equation}
 x_1^{18}+x_2^{18}+x_3^{18}+x_4^{3}+x_5^{2}-18 \psi x_1 x_2 x_3 x_4 x_5 -3 \phi x_1^6 x_2^6 x_3^6 \,,\label{18sub}
\end{equation}
i.e. all except two of the 272 complex structure moduli which correspond to monomials in the general degree 18 Calabi-Yau hypersurface equation have been set to zero. As was discussed in~\cite{Candelas:1994hw}, eq.~\eqref{18sub} is invariant under a global $\Gamma = \mathbb{Z}_6 \times\mathbb{Z}_{18}$ symmetry. This symmetry is used in the Greene-Plesser construction~\cite{Greene:1990ud} to construct the mirror Calabi-Yau which in this case has $h^{1,1}=272$ and $h^{2,1}=2$. Furthermore, the moduli $\psi$ and $\phi$ in eq.~\eqref{18sub} describe the two complex structure moduli of this mirror manifold. As was pointed out in~\cite{Candelas:2000fq}, the periods of the mirror agree with those of $\Pex$ at the $\Gamma$ symmetric point. Also,~\cite{Candelas:2000fq} shows that the complete set of $h^{2,1}$ complex structure moduli can be divided into a $\Gamma$-invariant subspace and its complement. The moduli with trivial transformation are exactly those that do not vanish at the $\Gamma$ symmetric point, in this case $\psi$ and $\phi$.

To make use of the agreement of the prepotential for the complex structure sector of $\Pex$ and its mirror in the large complex structure limit, it is useful to introduce the complex coordinates $U_1$ and $U_2$ that are related to $\psi$ and $\phi$ as~\cite{Candelas:1994hw}
\begin{align}
\begin{aligned}
 X_1 =-\frac{1}{q_1}&\left(1+312 q_1 + 2 q_2 +10260 q_1^2 - 540q_1 q_2 -q_2^2 \right.\\ &- \left. 901120 q_1^3 +120420 q_1^2 q_2 + 20 q_2^3 +\dots\right)\,,\\
 X_2 =-\frac{1}{q_2}&\left(1+180 q_1-6 q_2+11610 q_1^2+180 q_1 q_2 +27 q_2^2\right.\\ &+ \left. 514680 q_1^3-150120 q_1^2 q_2-5040 q_1 q_2^2-164 q_2^3 +\dots\right)\,,\label{XjUjrel}
\end{aligned}
\end{align}
up to third order in the $q_j \equiv e^{2\pi\, i \, U_j}$ with the large complex structure coordinates
\begin{equation}
 X_1 =\frac{(18\psi)^6}{3\phi}\qquad\qquad \text{and}\qquad \qquad X_2 = (3\phi)^3\,.
\end{equation}
The large complex structure limit corresponds to $X_j \to \infty$ which is equivalent to $\text{Im}(U_j)\to \infty$ as can be seen from eq.~\eqref{XjUjrel}.

There are two conifold singularities given by the equations~\cite{Candelas:1994hw}
\begin{eqnarray}
\begin{aligned}
 &\text{CF}_1:\,  (26244 \psi^6 + \phi)^3 =1 \qquad &\Leftrightarrow& \qquad \frac{X_2}{27}\left(\frac{X_1}{432}+1 \right)^3=1\,,\\
 &\text{CF}_2:\,  \phi^3 =1 \qquad &\Leftrightarrow& \qquad \frac{X_2}{27} = 1\,.\label{CF12}
\end{aligned}
\end{eqnarray}

Let us come back to the problem of finding supersymmetric extrema by solving eq.~\eqref{DiWall}. As was noted in~\cite{Giryavets:2003vd,Denef:2004dm}, to find an extremum it is sufficient to turn on fluxes only along the six $\Gamma$-invariant three-cycles, i.e.
\begin{equation}
 f=(f_{1_1},f_{1_2},f_{1_3},f_{2_1},f_{2_2},f_{2_3},0,...,0)\quad \text{and} \quad h=(h_{1_1},h_{1_2},h_{1_3},h_{2_1},h_{2_2},h_{2_3},0,...,0)\,,\label{6flux}
\end{equation}
having set to zero all the components along the $b_3-6$ non-invariant three-cycles. It is then possible to achieve $D_a W = 0$ for \textit{all} 272 complex structure moduli,\footnote{Note that orientifolding will project out some of the 272 complex structure moduli. Since the exact number of projected out directions depends on the position of the O-plane we stick to the upper bound of 272 for a general treatment.} and hence to find a minimum of the positive definite tree-level no-scale scalar potential eq.~\eqref{V4Dnoscale}.
This is due to the fact that, for this $\Gamma$ invariant flux, the symmetry $\Gamma$ is realized at the level of the four-dimensional effective action.
Note that the restriction to flux on the $\Gamma$ invariant cycles is purely for simplicity, as the analysis of the complete 272 dimensional complex structure moduli space is practically extremely challenging.

Let us explain more detailed why the flux vector in eq.~\eqref{6flux} generically provides a stable minimum of \textit{all} 272 complex structure moduli~\cite{Giryavets:2003vd,Denef:2004dm}. We first consider $D_{\tilde U_a} W_0 = 0$, where $\tilde U_a$ for $a=3,\dots,272$ denote the non-trivially transforming moduli under $\Gamma = \mathbb{Z}_{6}\times \mathbb{Z}_{18}$.
In the large complex structure limit, the prepotential $\mathcal{G}$ is a polynomial function of all $h^{2,1}$ complex structure moduli that has to transform trivially under $\Gamma$, since if it would not, $\Gamma$ could be used to fix the non-trivially transforming moduli.\footnote{$\mathcal{G}$ completely determines the moduli space of the (before orientifolding) $\mathcal{N}=2$ moduli space. If it would not be invariant the complex structure moduli space would have been reduced, i.e. some flat directions lifted but this does not happen just because there exists a $\Gamma$ symmetric point.} Hence, the non-trivially transforming $\tilde{U}_a$ have to appear at least quadratic in $\mathcal{G}$ in order to represent a $\Gamma$ invariant contribution to $\mathcal{G}$.
This information, together with having switched on flux only along the $\Gamma$ invariant directions, is sufficient to show
\begin{equation}
 W_{0,\tilde U_a} = K_{\tilde U_a} = 0 \quad \text{at} \quad \tilde U_a = 0 \quad \text{for} \quad a=3,\dots,272\,,
\end{equation}
since $W_{0,\tilde U_a}$ is a polynomial function which is at least linear in the $\tilde U_a$, see eq.~\eqref{W0period} and $K_{\tilde U_a}$ is a rational function which is at least linear in the numerator in the $\tilde U_a$, see eq.~\eqref{Kaehlercs}. Hence, $D_{\tilde U_a} W_0 = W_{0,\tilde U_a} + K_{\tilde U_a} W_0 =0$ at $\tilde U_a = 0$ for $a=3,\dots,272$.
This reduces the full set of conditions $D_aW=0$ $\forall a$ to the three equations
\begin{equation}
 D_{I}W |_{\tilde U_a = 0}=0 \qquad \mbox{for } \,\, I=\tau,U_1,U_2\,. \label{TOSOLVE}
\end{equation}
This is equivalent to set $\tilde U_a = 0$ from the beginning and study the stabilization problem for the reduced case with two complex structure moduli, as we do in the following.

In~\cite{Candelas:1994hw}, the prepotential $\mathcal{G}$ for the two complex structure moduli $U_1=\omega_1/\omega_0$ and $U_2=\omega_2/\omega_0$ was derived via mirror symmetry in the large complex structure limit to be
\begin{equation}
 \mathcal{G}(\omega_0,\omega_1,\omega_2) = \xi  \omega _0^2+\frac{17 \omega _0 \omega _1}{4}+\frac{3 \omega _0 \omega _2}{2}+ \frac{9 \omega _1^2}{4}+ \frac{3 \omega _1 \omega _2}{2}-\frac{9 \omega _1^3+9 \omega _1^2 \omega _2+3 \omega _1 \omega _2^2}{6 \omega _0}\,,\label{prepotex}
\end{equation}
with $\xi=\frac{\zeta(3)\chi}{2(2\pi\,i)^3}\simeq -1.30843\,i$ determined by the Euler number $\chi$ of the Calabi-Yau.

Eq.~\eqref{prepotex} receives instanton corrections which are given as
\begin{align}
 \mathcal{G}_{\text{inst.}}(q_1,q_2)=  \frac{1}{(2\pi\,i)^3}&\left( 540 q_1+\frac{1215 q_1^2}{2}+560 q_1^3+3 q_2-1080 q_1 q_2+143370 q_1^2 q_2 \right.\notag\\ & \left.-\frac{45 q_2^2}{2}+2700 q_1 q_2^2+\frac{244 q_2^3}{9} + \dots \right)\,,
\label{Ginst}
\end{align}
with $q_a = \exp{(2\pi\,i\,U_a)}$ and we have set $\omega_0=1$. The dots in eq.~\eqref{Ginst} denote higher powers in the $q_a$ which are suppressed in the large complex structure limit $u_a = \text{Im}(U_a) \gtrsim 1$. We define the large complex structure limit via
\begin{equation}
 \frac{|\mathcal{G}_{\text{inst.}}|}{|\mathcal{G}|} \leq \epsilon_{LCS}\,,\qquad \frac{540 e^{-2\pi u_1}}{(2\pi)^3|\mathcal{G}|} \leq \epsilon_{LCS}\qquad \text{and} \quad \frac{3 e^{-2\pi u_2}}{(2\pi)^3|\mathcal{G}|} \leq \epsilon_{LCS}\,,\label{largecompstr}
\end{equation}
for small $\epsilon_{LCS}$. The two last condition in eq.~\eqref{largecompstr} are imposed to ensure that there are no cancellations between the terms in $\mathcal{G}_{\text{inst.}}$, i.e. the leading correction in $e^{-2\pi u_a}$ is actually small. Furthermore, to have a valid description of the complex structure moduli by $\mathcal{G}$ we have to ensure that we are not in the vicinity of the conifold points eq.~\eqref{CF12}, i.e.
\begin{equation}
 \left|\frac{X_2}{27}\left(\frac{X_1}{432}+1 \right)^3 - 1\right| \geq \epsilon_{CF} \qquad \text{and} \qquad \left|\frac{X_2}{27} - 1\right| \geq \epsilon_{CF}\,,\label{CFcriterium}
\end{equation}
with small $\epsilon_{CF}$.

\section{The polynomial homotopy continuation method} \label{paramotopy_sec}

We want to solve the non-linear eqs.~\eqref{TOSOLVE} derived from the prepotential eq.~\eqref{prepotex} for the 6 real variables $x_i = u_1,u_2,s,\nu_1,\nu_2$ and $\sigma$. The parameters of these equations are the 12 fluxes $f_1,f_2,h_1$ and $h_2$ in eq.~\eqref{6flux}. Though systems of non-linear equations are extremely difficult
to solve in general, if the non-linearity in the system is polynomial-like,
then the recently developed algebraic geometry methods can rescue the situation. In particular, we use the so-called numerical
polynomial homotopy continuation (NPHC) method \cite{SW:95} which finds
all the solutions of the given system of polynomial equations. This method has been used in various problems in particle theory and statistical mechanics in Refs.~\cite{Mehta:2009,Mehta:2009zv,Mehta:2011xs,Mehta:2011wj,Kastner:2011zz,Nerattini:2012pi,Maniatis:2012ex,Mehta:2012wk,Hughes:2012hg,Mehta:2012qr,Hauenstein:2012xs}.

\subsection{The algorithm} \label{paramotopy_alg_sec}
Here we briefly explain the NPHC method: for a system of polynomial equations,
$P(x)=0$, where $P(x)=(p_{1}(x),\dots,p_{m}(x))^{T}$ and $x=(x_{1},\dots,x_{m})^{T}$,
which is \textit{known to have isolated solutions}, the \textit{Classical B\'ezout Theorem} asserts that
for generic values of coefficients, the maximum number of solutions in $\mathbb{C}^{m}$
is $\prod_{i=1}^{m}d_{i}$. Here, $d_{i}$ is the degree of the $i$th
polynomial. This bound, the \emph{classical B\'ezout bound} (CBB), is
exact for generic values \cite{SW:95,Li:2003}
for details.

Based on the CBB, a \textit{homotopy} can be constructed as
\begin{equation}
H(x,t)=\gamma(1-t)Q(x)+t\; P(x),
\end{equation}
where $\gamma$ is a generic complex number and $t\in[0,1)$. $Q(x)=(q_{1}(x),\dots,q_{m}(x))^{T}$
is a system of polynomial equations with the following properties:
\begin{enumerate}
\item the solutions of $Q(x)=H(x,0)=0$ are known or can be easily obtained.
$Q(x)$ is called the \textit{start system} and the solutions are
called the \textit{start solutions},
\item the number of solutions of $Q(x)=H(x,0)=0$ is equal to the CBB for
$P(x)=0$,
\item the solution set of $H(x,t)=0$ for $0\le t\le1$ consists of a finite
number of smooth paths, called homotopy paths, each parameterized by
$t\in[0,1)$, and
\item every isolated solution of $H(x,1)=P(x)=0$ can be reached by some
path originating at a solution of $H(x,0)=Q(x)=0$.
\end{enumerate}
The start system $Q(x)=0$ can for example be taken to be
\begin{equation}
Q(x)=\left(\begin{array}{c}
x_{1}^{d_{1}}-1\\
\vdots\\
x_{m}^{d_{m}}-1
\end{array}\right)=0,\label{eq:Total_Degree_Homotopy}
\end{equation}
where $d_{i}$ is the degree of the $i^{th}$ polynomial of the original
system $P(x)=0$. Eq.~(\ref{eq:Total_Degree_Homotopy}) is easy to
solve and guarantees that the total number of start solutions is $\prod_{i=1}^{m}d_{i}$,
all of which are non-singular.

We can then track all the paths corresponding to each solution of
$Q(x)=0$ from $t=0$ to $t=1$. The paths which reach $P(x)=0=H(x,1)$ are the solutions of $P(x)=0$. By implementing
an efficient path tracker algorithm, all isolated solutions of a system of multivariate
polynomials system can be obtained because it is shown \cite{SW:95} that for a generic $\gamma$, there are no singularities (i.e., paths do not cross
each other) for $t\in[0,1)$. In this respect, the NPHC method has a great advantage
over all other known methods for finding stationary points.

There are several sophisticated numerical packages well-equipped with
path trackers such as Bertini\cite{BHSW06}, PHCpack~\cite{Ver:99},
PHoM~\cite{GKKTFM:04} and HOM4PS2 \cite{GLW:05,Li:2003}. We
mainly use Bertini to get the results in this
paper.

We mean by a solution a set of values of variables which satisfies the eqs.~\eqref{TOSOLVE} with tolerance $10^{-10}$. All the solutions
come with real and imaginary parts. A solution is a real solution
if the imaginary part of each of the variables is less than or equal
to the tolerance $10^{-6}$ (below which the number of real solutions
does not change, i.e., it is robust for the problem at hand). All these solutions can be
further refined to an \textit{arbitrary precision}.

The advantages of the homotopy based on the CBB are (1) the CBB is easy
to compute, and (2) the start system based on the CBB can be solved quickly.
The drawback of it is that the CBB does not take the
sparsity of the system into account: systems arising in practice
have far fewer solutions than the CBB, so a large portion
of the computational effort is wasted.

Hence, one can also use homotopies based on tighter upper bounds. For example, one can compute the so-called 2-Homogeneous B\'ezout Bound or the Bernstein-Khovanskii-Kushnirenko bound~\cite{Bernstein75,Khovanski78,Kushnirenko76} which are tighter upper bounds. These two bounds were explained in Ref.~\cite{Mehta:2009,Mehta:2012wk}. We note that, as with the CBB, the 2HomBB and BKK bound are also generically sharp with respect to the family of polynomial systems under consideration.

There is yet another, rather more practical, way of solving a \textit{parametric} system which is called Cheater's homotopy \cite{li1989cheater,li1992nonlinear}: let us say we want to solve a parametric system, $\vec{f}(\vec{q}; \vec{x}) = \vec{0}$ where $\vec{x}$ are variables and $\vec{q}$ are parameters, in our case the fluxes. Now, it can be shown \cite{li1989cheater,li1992nonlinear} that the maximum number of complex solutions at any parameter point is the number of solutions at a generic parametric point. So our strategy is first to solve the system at a generic parameter point and then using the solutions at this point as the start solutions for the systems at all other parameter-points. This homotopy is called cheater's homotopy. A recently developed software, based on Bertini, called Paramotopy~\cite{paramotopytoappear}, precisely uses cheater's homotopy and goes over a huge number of parameter points in parallel. The package is publicly not available yet though the respective research group has kindly given access to the code for the purpose of the computation in this paper. We will publish the details on the cheater's homotopy and the package Paramotopy elsewhere~\cite{cheatershomotopytoappear}.

\subsection{The scan} \label{scan_param_sec}

We define a set of flux parameters on which we apply the algorithm described in the previous section~\ref{paramotopy_alg_sec}. Since we are only interested in supersymmetric flux vacua, we can make use of the ISD condition eq.~\eqref{ISDF3H3} and define a flux configuration via
\begin{equation}
 H_3 = \left( \begin{array}{c}
         h_1\\ h_2
        \end{array}
 \right) \qquad \quad \text{and} \qquad \quad F_3 = \left( \begin{array}{c}
         -h_2\\ h_1
        \end{array}
 \right)\,,\label{h1h2}
\end{equation}
with $h_1,h_2\in \mathbb{Z}^3$. Note that since we have two complex structure moduli we have initially $2\cdot2+2=6$ flux parameters for both $H_3$ and $F_3$ but the ISD condition eq.~\eqref{ISDF3H3} reduces this to the six parameters given in eq.~\eqref{h1h2}. Furthermore, the D3 tadpole eq.~\eqref{D3tad} becomes manifestly positive semi-definite, i.e.
\begin{equation}
 L=h_1^2 +h_2^2 \,.\label{tadpolesphere}
\end{equation}

To scan efficiently, we apply the paramotopy algorithm only to $SL(2,\mathbb{Z})$ inequivalent flux configurations. Note that a configuration of the form eq.~\eqref{h1h2} transforms as
\begin{equation}
  H_3' = \left( \begin{array}{c}
          h_1'\\ h_2'
        \end{array}
 \right) = \left( \begin{array}{c}
        d\,h_1-c\,h_2\\ d\,h_2+c\,h_1
        \end{array}
 \right)\qquad \text{and} \qquad  F_3' = \left( \begin{array}{c}
         f_1'\\ f_2'
        \end{array}
 \right)= \left( \begin{array}{c}
        b\,h_1-a\,h_2\\ b\,h_2+a\,h_1
        \end{array}\right)\,,\label{h1'h2'}
\end{equation}
under $SL(2,\mathbb{Z})$ transformations, eq.~\eqref{SL2Zgen}. For general $a,b,c$ and $d$, $F_3'$ in eq.~\eqref{h1'h2'} will not be of the form in eq.~\eqref{h1h2} but only for the 4 cases
\begin{equation}
 a=\pm1,\,b=0,\,c=0,\,d=\pm1 \qquad \text{and} \qquad  a=0,\,b=\pm1,\,c=\mp1,\,d=0\,.\label{abcdinsphere}
\end{equation}
This corresponds to the $SL(2,\mathbb{Z})$ equivalent flux configurations
\begin{equation}
 \left( \begin{array}{c}
         h_1\\ h_2
        \end{array}
 \right) \cong
 \left( \begin{array}{c}
         -h_1\\ -h_2
        \end{array}
 \right) \cong
 \left( \begin{array}{c}
         -h_2\\ h_1
        \end{array}
 \right) \cong
 \left( \begin{array}{c}
         h_2\\ -h_1
        \end{array}
 \right)\,.\label{4h1h2}
\end{equation}
Note that the two transformations on the LHS of eq.~\eqref{abcdinsphere} do not transform the dilaton $\tau'=\tau$ while the two transformations on the RHS act as $\tau'=-1/\tau$.

The number of $SL(2,\mathbb{Z})$ inequivalent flux configurations in a spherical region defined by a spherical constraint eq.~\eqref{tadpolesphere} can be estimated as $\pi^3/(4\,\Gamma[4]) (\sqrt{L})^6$, using the formula for the volume of the n-sphere $V_n(r)=\pi^{n/2}/\Gamma(n/2+1)\,r^n$. The factor $1/4$ accounts for the 4 equivalent configurations in eq.~\eqref{4h1h2}. If we had switched on more flux $n>6$ the number of lattice points grows very rapidly $\sim L^{n/2}$.

For our scan, we choose $L=34$ such that we scan over 52,329 parameter points (the above estimate yields 55,391). On the \textit{FermiLab cluster} using $100$ nodes each with $32$ cores (each core with 2.0 GHz cloak speed), the calculation time in total is around $75,000$ hours, with $60-100$ minutes per parameter point.

\subsection{Distribution of parameters} \label{distr_param_sec}

In this section, we want to discuss the distribution of the following parameters as results of the scan defined in the previous section~\ref{scan_param_sec}:
\begin{itemize}
 \item $u_1$ and $u_2$, to see how many points reach the large complex structure limit as defined in eq.~\eqref{largecompstr}.
 \item $\tau$, to identify regions of weak respectively strong coupling.
 \item The number of solutions for a given D3 tadpole $L$.
 \item $W_0$, the flux superpotential.
 \item The masses of the moduli $m^2$ and the gravitino mass $m^2_{3/2}$.
 \item The available fine-tuning $\Delta \Lambda$ of the cosmological constant $\Lambda$.
 \item The amount of flux vacua for which a dS vacuum can be constructed via K\"ahler uplifting in~\ref{deSitter_param_sec}.
\end{itemize}

For the 52,329 parameter points, we find a total of 500,865 solutions to the eqs.~\eqref{TOSOLVE}. This corresponds to an average of 9.5 solutions per parameter point. For 1,270 parameter points we do not find a solution. Many of the solutions are unphysical and hence have to be sorted out: A subset of 271,825  fulfill the criterion of a physical string coupling $g_s > 0$ and only a subset of 24,882 respective 15,392 is in accordance with the large complex structure criterion eq.~\eqref{largecompstr} for $\epsilon_{LCS} = 10^{-1}$ respective $\epsilon_{LCS} = 10^{-2}$. Of these none have to be sorted out because they are in the vicinity of the conifold singularities eq.~\eqref{CFcriterium} having chosen $\epsilon_{CF} = 10^{-2}$. Due to the strong suppression of the large complex structure limit in the general solution space of eq.~\eqref{TOSOLVE} the minimal flux method has the advantage of directly searching for solutions in this region.

\begin{figure}[t!]
\centering
\includegraphics[width= \linewidth]{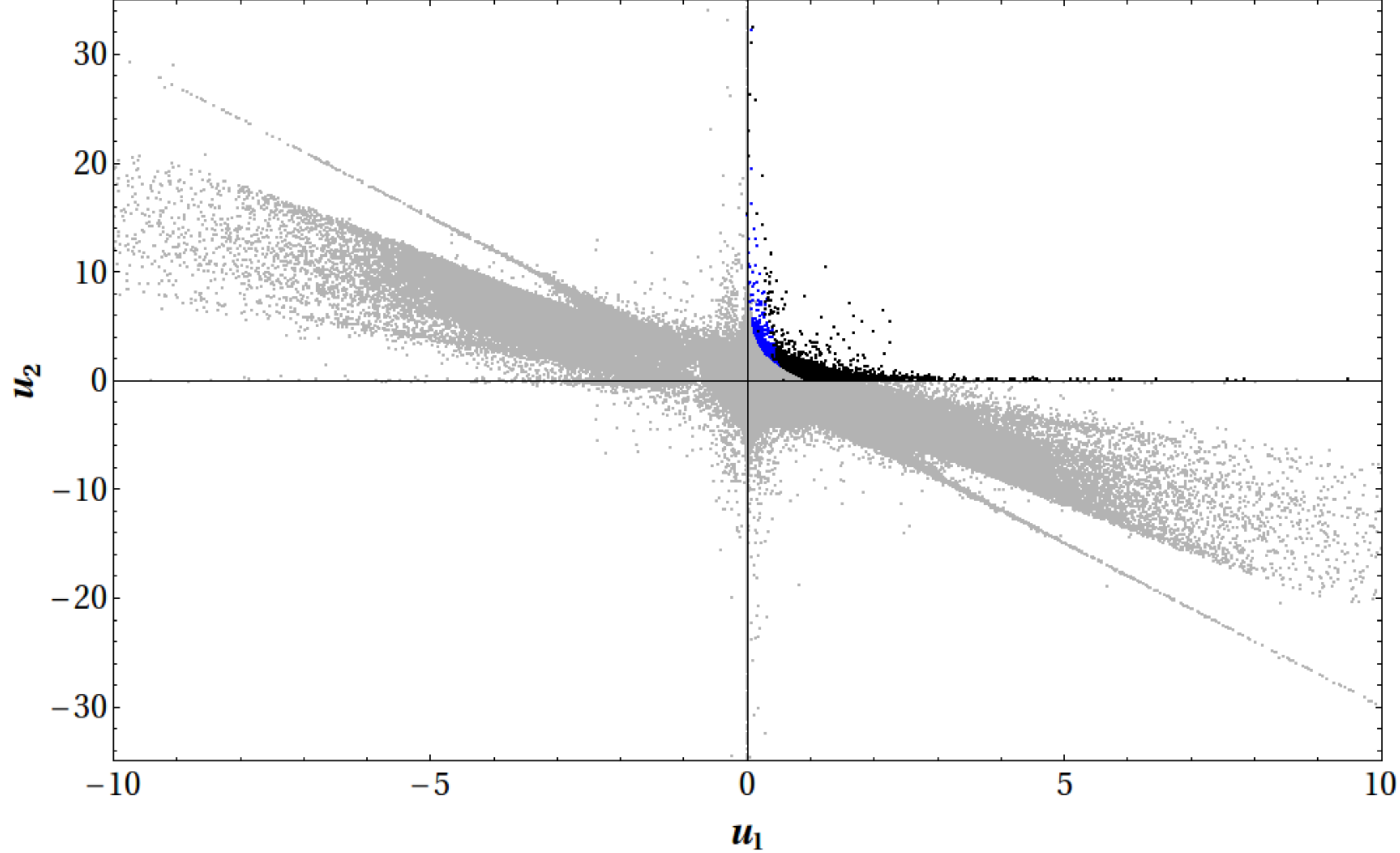}
\caption{Distribution of $u_1$ and $u_2$ for the complete set of solutions to eqs.~\eqref{TOSOLVE} (gray) and for the physical solutions fulfilling the criterion for the validity of the large complex structure limit (blue for $\epsilon_{LCS} = 10^{-1}$ and black for $\epsilon_{LCS} = 10^{-2}$), eq.~\eqref{largecompstr}.}
\label{LCS_param_fig}
\end{figure}

\hskip -5mm
\begin{minipage}{0.51\linewidth}
For the distribution of the dilaton, we can use $SL(2,\mathbb{Z})$, to transform each solution to the fundamental domain
\begin{equation}
 -\frac{1}{2} \leq \text{Re}(\tau) \leq \frac{1}{2} \qquad \text{ and } \qquad |\tau|>1\,,\label{funddomain}
\end{equation}
via the successive transformations
\begin{equation}
 \tau' = \tau+b\,, \quad G_3' = G_3\,,\label{S+btrans}
\end{equation}
i.e. $a=1,\,b\in\mathbb{Z},\,c=0,\,d=1$ and
\begin{equation}
 \tau' = -1/\tau\,, \quad G_3' = G_3/\tau\,,\label{-1Strans}
\end{equation}
i.e. $a=0,\,b=-1,\,c=1,\,d=0$.

We show the distribution of the obtained values for $\tau=\sigma + i\, s$ in Figure~\ref{dilatondistr_param_fig}. We see that the the strongly coupled region $s = 1/g_s \sim 1$ is preferred and large values of $s>10$ are obtained for a fraction of $5\%$.
\end{minipage}
\hskip 5mm
\begin{minipage}{0.45\linewidth}
\begin{center}
\includegraphics[width= \linewidth]{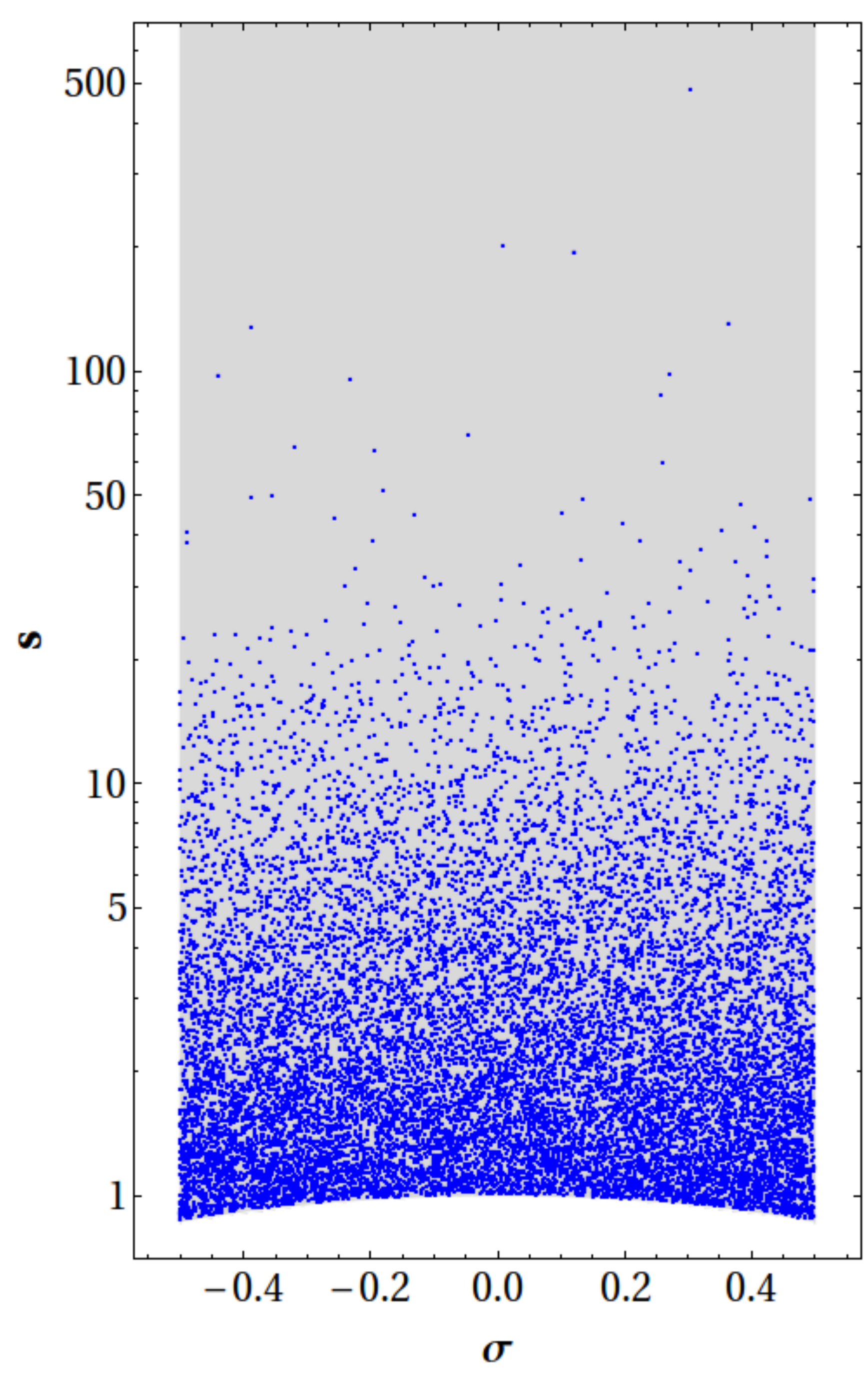}
\captionof{figure}{Distribution of $\tau$ for the paramotopy scan with $\epsilon_{LCS} = 10^{-2}$.}
\label{dilatondistr_param_fig}
\end{center}
\end{minipage}

\begin{figure}[t!]
\centering
\includegraphics[width= 0.48\linewidth]{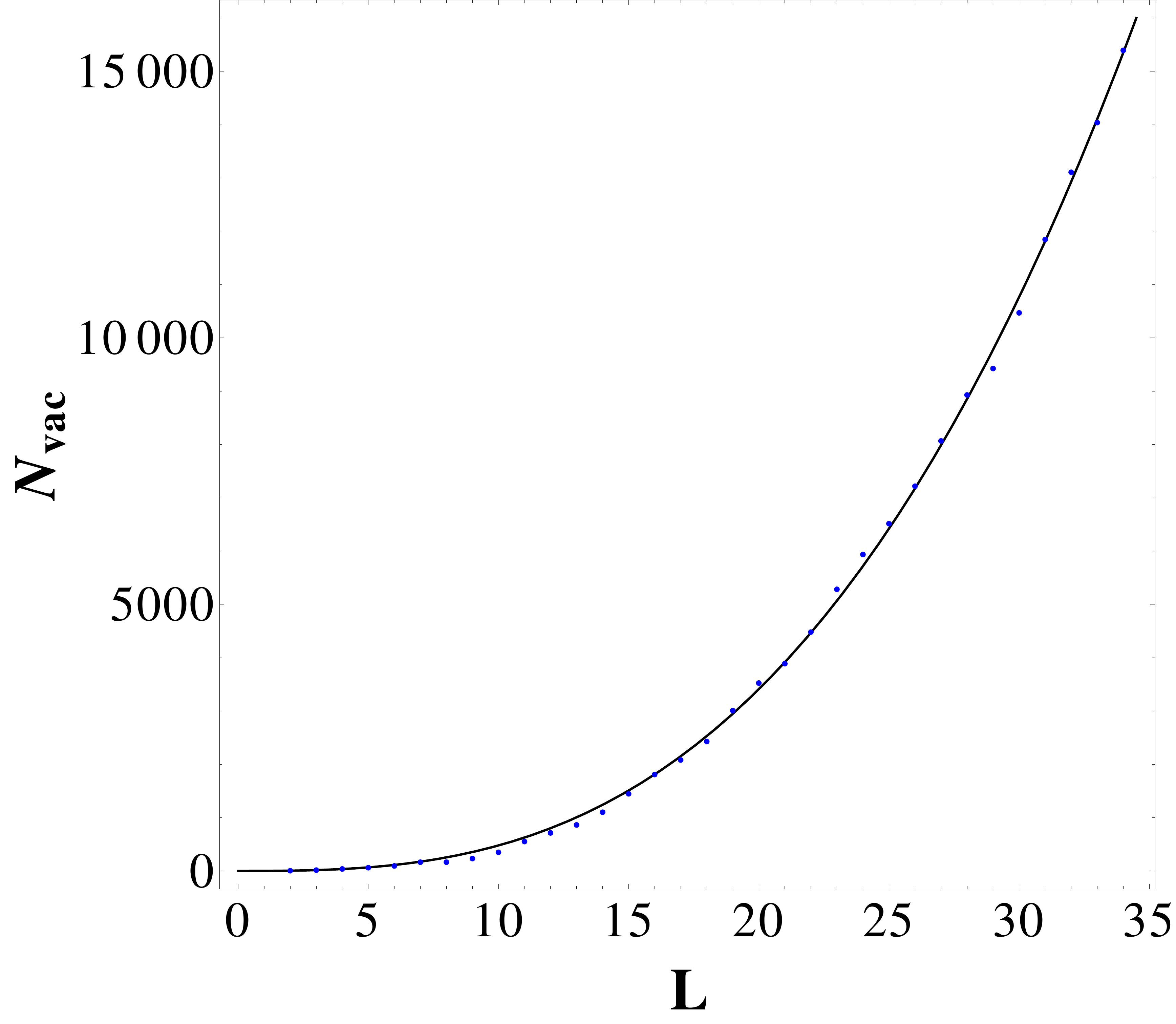}
\hskip 3mm
\includegraphics[width= 0.48\linewidth]{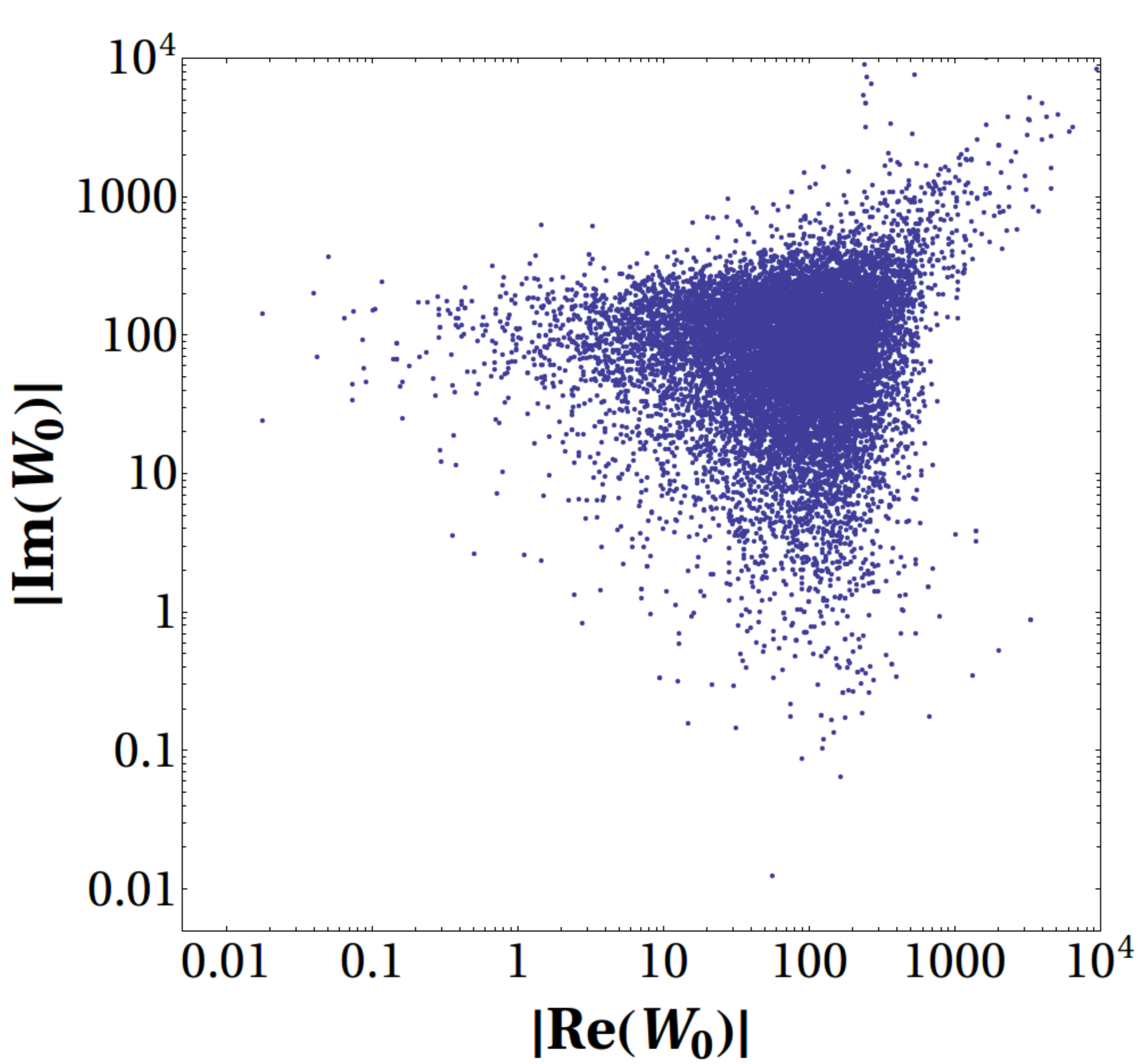}
\caption{The number of vacua $N_{vac}$ with $h^2<L$ (left) and the logarithmic distribution of the flux superpotential $W_0$ (right) in the large complex structure limit with $\epsilon_{LCS} = 10^{-2}$.}
\label{TadpoleW0_param_fig}
\end{figure}

The number of vacua of $X_3$ in the large complex structure limit for a given D3 tadpole $L$ was estimated in~\cite{Denef:2004dm} as~\footnote{Note that $N_{vac} \sim L^6$ in~\cite{Denef:2004dm} which is due to the fact that 12 independent fluxes have been switched on while we effectively switch on 6 independent fluxes, see eq.~\eqref{h1h2}.}
\begin{equation}
N_{vac}  = \frac{(2\pi L)^3}{3!} \int \det (-\mathcal{R} - \boldsymbol{1}\cdot \omega)\,,\label{NvacDenefDouglas}
\end{equation}
with K\"ahler form $\omega$ and the curvature two-form $\mathcal{R}$ of the moduli space. The integral in eq.~\eqref{NvacDenefDouglas} was estimated in~\cite{Denef:2004dm} to be be $1/1296$, using the $\Gamma$ symmetry of the moduli space such that
\begin{equation}
 N_{vac} \simeq 0.03\, L^3\,.
\end{equation}
Since paramotopy allows us to find all solutions for a given flux configuration we can not only check the $L$ dependence of eq.~\eqref{NvacDenefDouglas} but also the normalization. This depends on the value chosen for $\epsilon_{LCS}$, i.e. a greater $\epsilon_{LCS}$ will yield a larger normalization factor. Fitting the number of solutions with $h^2\leq L$ in the large complex structure limit to the tadpole $L$ we find
\begin{eqnarray}
&N_{vac} \simeq (0.52\pm 0.04)\, L^{2.92\pm 0.03} \qquad &\text{for} \qquad \epsilon_{LCS} = 10^{-2}\,,\label{Nvacfit1}\\
&N_{vac} \simeq (0.88\pm 0.06)\, L^{2.91\pm 0.03} \qquad &\text{for} \qquad \epsilon_{LCS} = 10^{-1}\,.
\end{eqnarray}
The dependence of $N_{vac}$ on $L$ and the fit in eq.~\eqref{Nvacfit1} are shown in Figure~\ref{TadpoleW0_param_fig}. Considering the very general arguments that are used to derive the estimate eq.~\eqref{NvacDenefDouglas}, the agreement within an order of magnitude with the factual number of vacua strongly confirms the statistical analysis of~\cite{Ashok:2003gk,Denef:2004ze}. In the following, we set $\epsilon_{LCS} = 10^{-2}$.

The distribution of the flux superpotential is shown in Figure~\ref{TadpoleW0_param_fig}. We find that for most vacua $\mathcal{O}(10^1-10^3)$ values are preferred. To calculate the masses of the moduli we have to know the value of the volume $\Vol$ of $X_3$ which enters via the K\"ahler potential of the K\"ahler moduli given in eq.~\eqref{KaehlerK}. Note that we have not specified the stabilization mechanism for the K\"ahler moduli and hence have no information about the value of $\Vol$. For the KKLT and K\"ahler uplifting scenarios the volume is typically stabilized at $\mathcal{O}(10^2-10^4)$ while for the LVS it is $\mathcal{O}(10^6-10^{15})$. Hence, we can only calculate the physical masses $m$ up to factors of $\Vol^{-1}$, i.e.
\begin{equation}
 m = \frac{m_{\text{cs}}}{\Vol}\,,
\end{equation}
where $m_{\text{cs}}$ is the mass calculated from the effective theory of the complex structure moduli only, i.e.$K=K_{\text{cs}}$ and $W=W_0$.

\begin{figure}[t!]
\centering
\includegraphics[width= \linewidth]{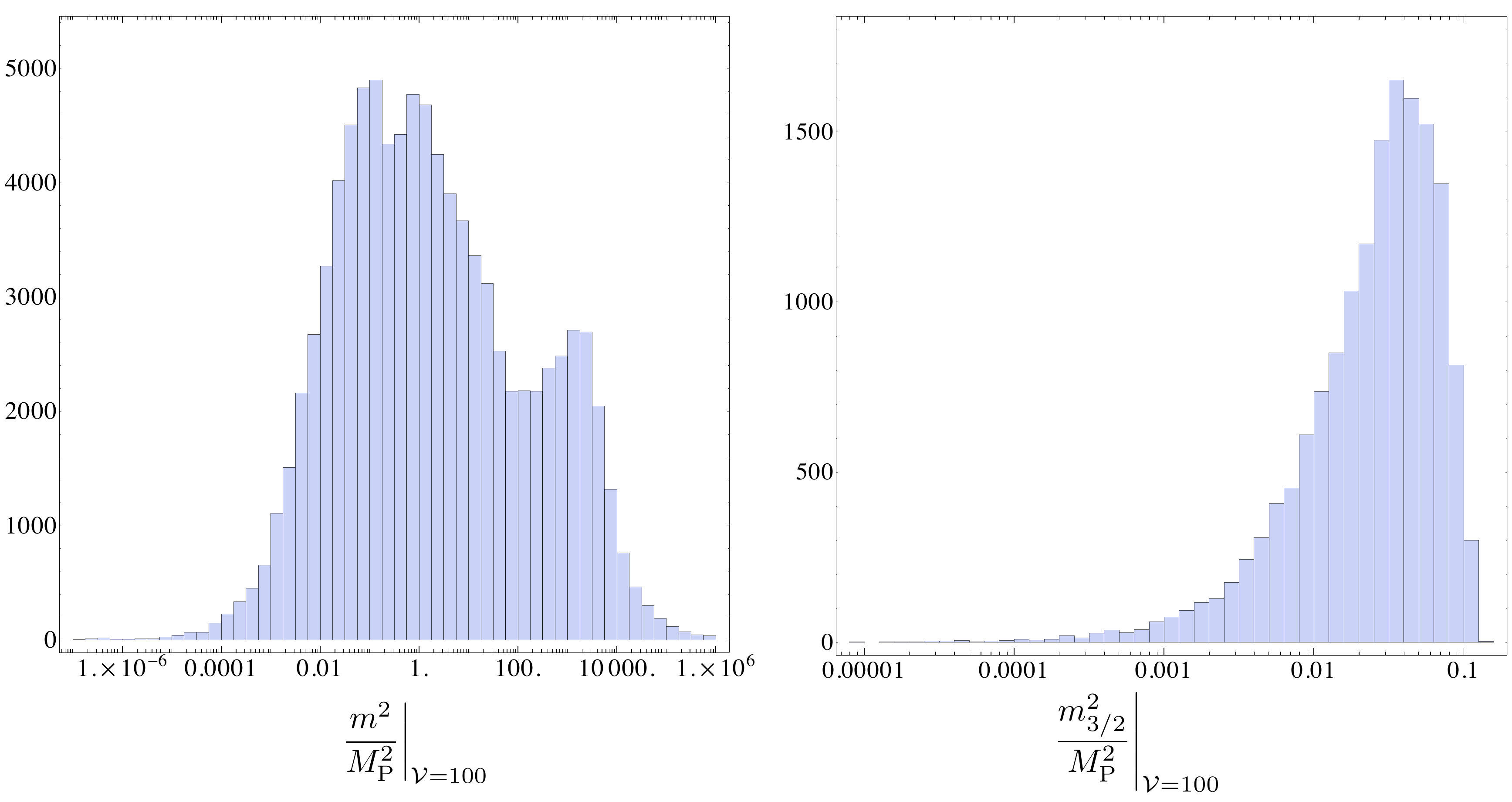}
\caption{Masses $\left. \frac{m^2}{M_{\text{P}}^2}\right|_{\Vol=100}$ of the moduli, and the gravitino mass $\left. \frac{m_{3/2}^2}{M_{\text{P}}^2}\right|_{\Vol=100}$, for a Calabi-Yau volume $\Vol=100$. For different Calabi-Yau volumes the masses scale as $\left. \frac{m^2}{M_{\text{P}}^2}\right|_{\Vol=100} \left(\frac{100}{\Vol} \right)^2$ for the moduli $u_1,u_2,s,\nu_1,\nu_2,\sigma$ (left) and for the gravitino mass as $\left. \frac{m_{3/2}^2}{M_{\text{P}}^2}\right|_{\Vol=100} \left(\frac{100}{\Vol} \right)^2$ (right). The left plot includes all eigenvalues of the Hessian, i.e. 6 values per parameter point.}
\label{modmassm32_param_fig}
\end{figure}

The distribution of the physical moduli masses $m^2$ in terms of $m^2_{\text{cs}}$, i.e. the eigenvalues of the Hessian $\partial_a \partial_b V$ of the no-scale potential eq.~\eqref{V4Dnoscale} for $a,b=u_1,u_2,s,\nu_1,\nu_2,\sigma$ is shown in Figure~\ref{modmassm32_param_fig} as well as the gravitino mass $m_{3/2}^2$ in terms of the quantity
\begin{equation}
 \gravm^2 \equiv m_{3/2}^2\, \Vol^2 =  e^{K_{\text{cs}}}|W_0|^2\,.
\end{equation}
This quantity $\gravm^2$ governs the scale of the typical AdS cosmological constant induced by the flux superpotential ignoring the contributions from the K\"ahler moduli sector.

The distribution of $m_{3/2}^2$ is peaking at $\langle m_{3/2}^2 \rangle = 3.5\times 10^{-2}\cdot(\frac{100}{\Vol})^{2}$ with standard deviation $3\times 10^{-2}\cdot(\frac{100}{\Vol})^{2}$. The complex structure moduli and the dilaton are stabilized at $m^2 \sim \mathcal{O}(10^{-3}-10^2) (\frac{100}{\Vol})^{2}$. These ranges for the moduli and gravitino masses are compatible with the values obtained for a single explicit flux choice in the same construction~\cite{Rummel:2011cd,Louis:2012nb}.

The AdS respective dS cosmological constant before tuning is up to $\mathcal{O}(1)$ factors estimated to be
\begin{equation}
 \Lambda \sim \frac{m_{3/2}^2}{\Vol} = \frac{\gravm^2}{\Vol^3}\,,\label{Lambdam32}
\end{equation}
in the LVS respective K\"ahler uplifting scenarios. In particular, the tunability of $\gravm$ by three-form flux directly translates into the tunability of the cosmological constant via
\begin{equation}
 \frac{\Delta \Lambda}{\Lambda} \sim 2\,\frac{\Delta \gravm}{\gravm}\,.\label{cctuning}
\end{equation}
Note that the RHS of eq.~\eqref{cctuning} is independent of the volume $\Vol$, i.e. fine tuning of $\gravm$ only has a tiny effect on the VEVs of the K\"ahler moduli. 

Since the polynomial homotopy continuation method allows us to calculate all solutions for a given tadpole $L$ we can estimate $\Delta \Lambda / \Lambda$ by determining the average spacing $\Delta \gravm$ for all values of $\gravm$ that are to be found in a $\sigma$-interval around $\langle \gravm \rangle$. Since the number of vacua is given as a power-law in $L$, with the exponent linear in the number of flux carrying complex structure moduli $h^{2,1}_{\text{eff}}$, we expect $\Delta \gravm / \gravm$ to be of the form
\begin{equation}
 \frac{\Delta \gravm}{\gravm} \sim \frac{C}{L^{a\,( h^{2,1}_{\text{eff}}+1)}}\,,\label{gravspacingfit}
\end{equation}
with $C,a >0$. We can determine these parameters by fitting the LHS of eq.~\eqref{gravspacingfit} as a function of $L$ for $h^{2,1}_{\text{eff}}=2$. Choosing a 3-$\sigma$ interval~\footnote{There is only a week dependence on the width of the interval. For 5-$\sigma$ the difference in $\Delta \gravm / \gravm$ compared to 3-$\sigma$ is less than 1\%.} around $\langle \gravm \rangle$ we find the available tuning for the cosmological constant to be
\begin{equation}
 \frac{\Delta \Lambda}{\Lambda} \simeq (5.1 \pm 0.3)\,L^{- (0.93\pm 0.006)\, ( h^{2,1}_{\text{eff}}+1)}\,, \label{DeltaLambdafit}
\end{equation}
where we have included the statistical errors of the fit parameters $C$ and $a$.

Let us assume that eq.~\eqref{DeltaLambdafit} is valid and $\langle \gravm \rangle \sim \mathcal{O}(10)$ also for larger values of $L\sim \mathcal{O}(10^3)$ and larger values of $h^{2,1}_{\text{eff}} \sim \mathcal{O}(10^1-10^2)$.~\footnote{This assumption is reasonable when the prepotential $\mathcal{G}$ is of the same structure as eq.~\eqref{prepotex}, i.e. we are considering the large complex structure limit away from e.g. conifold singularities via a mirror construction. It may be interesting to consider such examples with $h^{2,1}_{\text{eff}} =h^{1,1}> 2$, e.g. by choosing random pre-factors in a general polynomial prepotential of degree 2 in the $\omega_i$.} Then, we can extrapolate the values of the cosmological constant eq.~\eqref{Lambdam32} and its tunability to more realistic scenarios, see Table~\ref{gravitinomasscc_tab}.

\begin{table}[ht!]
\centering
  \begin{tabular}{|c|c||c|}
  \hline
  $h^{2,1}_{\text{eff}}$ & $L$ & $\Delta \Lambda / \Lambda$\\
  \hline
  \hline
  $2$ & $34$ & $7\cdot10^{-3} \pm 5\cdot10^{-4}$\\
  \hline
  $2$ & $500$ & $5\cdot 10^{-5} \pm 4\cdot10^{-6}$\\
  \hline
  $40$ & $34$ & $5\cdot10^{-57} \pm 4\cdot10^{-57}$\\
  \hline
  $40$ & $500$ & $2\cdot10^{-100} \pm 2\cdot10^{-100}$\\
  \hline
  \end{tabular}
  \caption{The tunability $\Delta \Lambda / \Lambda$ of the cosmological constant for different values of $h^{2,1}_{\text{eff}}$ and $L$ with statistical errors propagated from eq.~\eqref{DeltaLambdafit}. The untuned values of the cosmological constant are estimated via eq.~\eqref{Lambdam32} to be $\mathcal{O}(10^{-4} - 10^{-22})$ in units of $M_{\text{P}}^4$ for $\Vol$ of $\mathcal{O}(10^2-10^8)$. The first row of this table is directly calculated from our dataset while the last three entries are obtained as an extrapolation via eq.~\eqref{DeltaLambdafit}.}
  \label{gravitinomasscc_tab}
\end{table}

To tune the cosmological constant to the accuracy given in Table~\ref{gravitinomasscc_tab}, one has to make the assumption that every supersymmetric flux vacuum has no tachyonic directions after uplifting and stabilizing the K\"ahler moduli. Especially, for large values of $h^{1,1}$ there could be strong suppressions of tachyonic free configurations~\cite{Marsh:2011aa,Chen:2011ac,Bachlechner:2012at}. In the following section, we will determine how many de Sitter vacua can be constructed from our dataset on $X_3$ via the method of K\"ahler uplifting.

\subsection{de Sitter vacua via K\"ahler uplifting} \label{deSitter_param_sec}

The two K\"ahler moduli $T_1$ and $T_2$ of $X_3$ can be stabilized in a dS minimum by K\"ahler uplifting. A globally consistent D7 brane and gauge flux setup that realizes such a dS vacuum has been presented in~\cite{Louis:2012nb}. The K\"ahler potential of $T_1$ and $T_2$ is given as
\begin{equation}
 K = -2 \log \left[ \frac{1}{\sqrt{12}} \left((T_1 +  \bar T_1) + \frac{1}{3} (T_2 + \bar T_2) \right)^{3/2} - \frac{1}{18} (T_2 + \bar T_2)^{3/2} + \frac{1}{2}\hat\xi(\tau,\bar \tau) \right]\,,
\end{equation}
with the leading order $\alpha'$ correction~\cite{Becker:2002nn}
\begin{equation}
 \hat\xi(\tau,\bar \tau)= - \frac{\zeta(3)\,\chi}{4 \,\sqrt{2}\, (2 \pi)^3}   \, (-i\,(\tau -\bar \tau))^{3/2}\,,\label{xihatgen}
\end{equation}
with Euler number $\chi = 2(2-272) = -540$. To apply the method of K\"ahler uplifting we need to balance the leading order $\alpha'$ correction to the K\"ahler potential with non-perturbative contributions to the superpotential. These originate from gaugino condensation of an $SU(24)$ and $SO(24)$ pure super Yang Mills from respectively 24 D7 branes wrapping the divisors corresponding to $T_1$ and $T_2$. The induced superpotential is
\begin{equation}
 W = W_0 + A_1\,e^{- \frac{2\pi}{24} T_1} + A_2\,e^{- \frac{2\pi}{22} T_2}\,.
\end{equation}
By switching on suitable gauge flux it can be shown~\cite{Louis:2012nb} that $A_1, A_2 \neq 0$. The induced D3 tadpole by this gauge flux and the geometric contributions from the D7 branes is $L=96-104$. The one-loop determinants $A_1$ and $A_2$ depend on the complex structure moduli, the dilaton and also potentially D7 brane moduli. The explicit dependence on these moduli is unknown, however for the purpose of K\"ahler moduli stabilization the values of $A_1$ and $A_2$ can be assumed be constant since complex structure moduli are stabilized at a higher scale. Choosing $A_1=A_2=1$, it was found numerically in~\cite{Louis:2012nb}, that the pairs of $W_0$ and $s$ that are suitable to realize a dS vacuum with small positive tree level vacuum energy~\footnote{In this case, small refers to how small we can tune $\langle V \rangle$ by choosing numerical values for $W_0$ and $s$ to a certain decimal place and is not related to the tuning of the cosmological constant.} lie on the curve
\begin{equation}
 W_0^{\text{dS}}(s) = 70.2\, s^{-2.35}\,\qquad \text{with} \qquad s\geq4\,.
\end{equation}

To parametrize our missing knowledge of the values of $A_1$ and $A_2$ we introduce the parameter $\Delta A$ and the scaling relations
\begin{equation}
 W_0 \to W_0 \cdot \Delta A\,,\qquad A_1 \to A_1\cdot \Delta A\,,\qquad A_2 \to A_2 \cdot \Delta A\,,
\end{equation}
under which the position of a minimum of the potential eq.~\eqref{V4D} is invariant since $V\to V\cdot \Delta A^2$. For a given uncertainty in the one-loop determinants $\Delta A^{-1} \leq A_i \leq \Delta A$ around $A_1 = A_2 = 1$ we can then define the criterion
\begin{equation}
 \frac{W_0^{\text{dS}}(s)}{\Delta A} \leq |W_0| \leq W_0^{\text{dS}}(s) \cdot \Delta A\, \qquad \text{and}\qquad s \geq 4\,,
\end{equation}
for a given data point $(s,|W_0|)$ to allow a dS vacuum via K\"ahler uplifting.

\begin{figure}[t!]
\centering
\includegraphics[width= 0.48\linewidth]{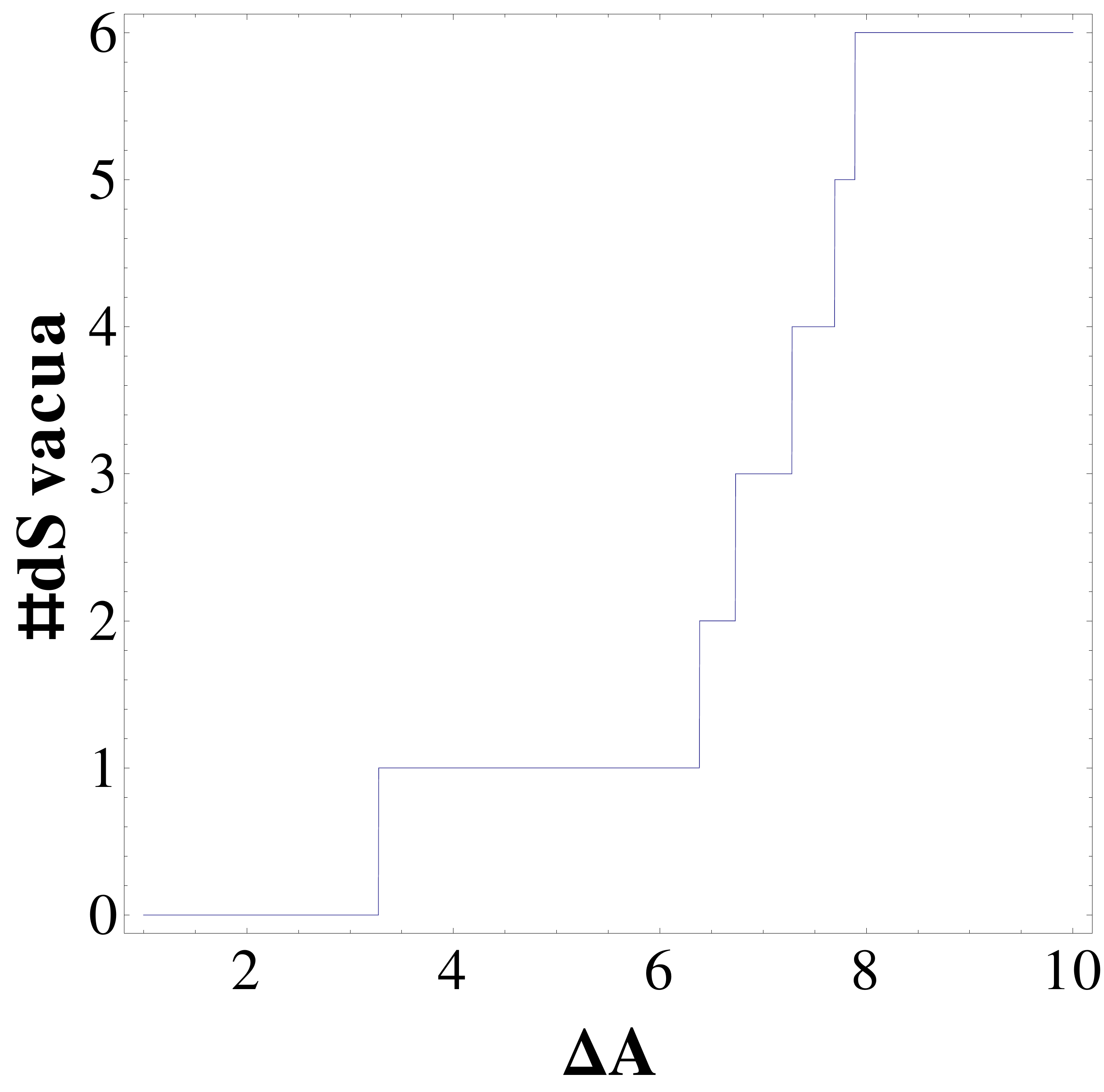}
\hskip 3mm
\includegraphics[width= 0.48\linewidth]{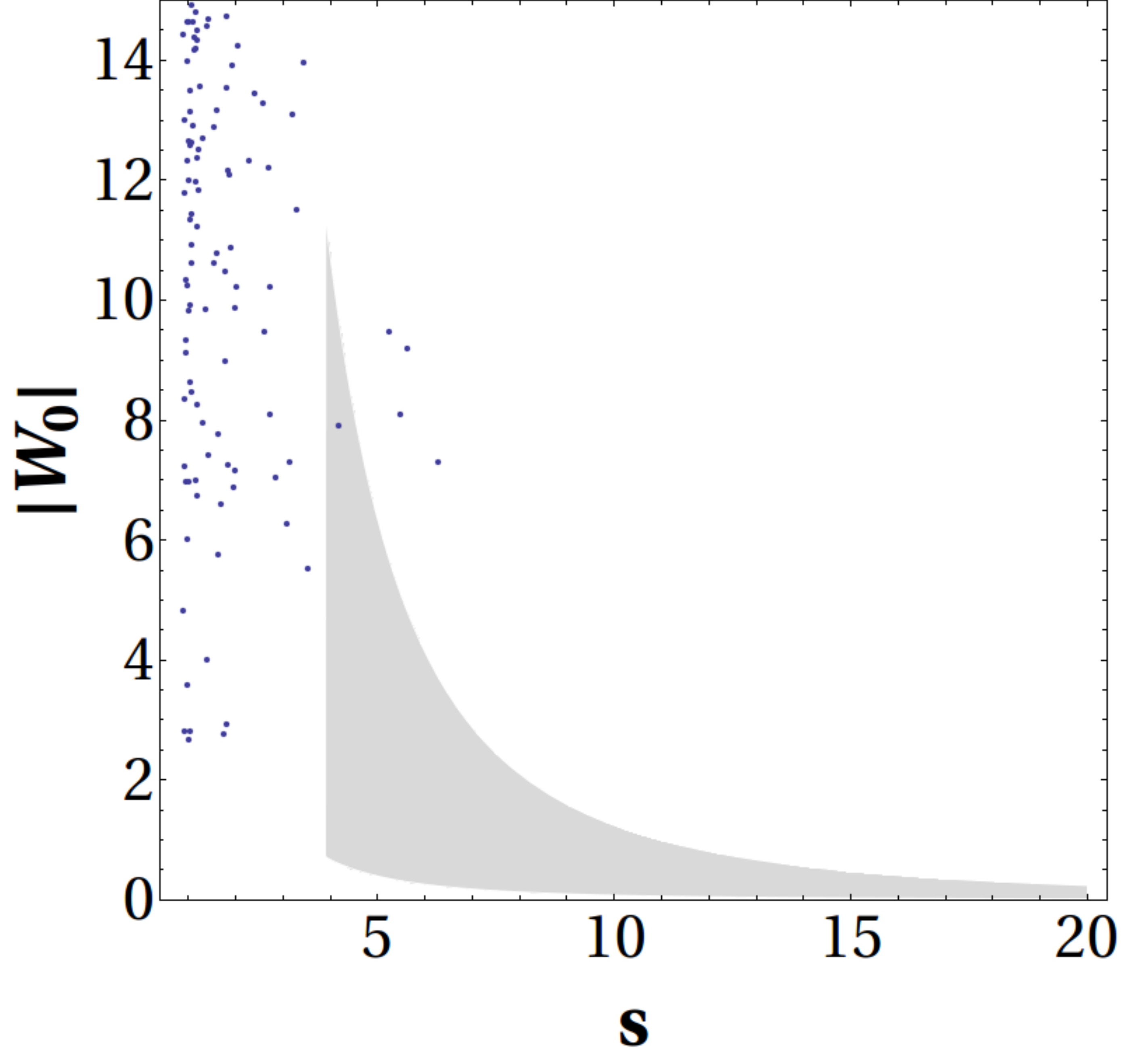}
\caption{The number of K\"ahler uplifted dS vacua as a function of $\Delta A$ (left) and data points $(s,|W_0|)$ (right). K\"ahler uplifting can be applied in the shaded region ($\Delta A =2$).}
\label{deSitter_param_fig}
\end{figure}

We show the number of K\"ahler uplifted dS vacua depending on $\Delta A$ in Figure~\ref{deSitter_param_fig}. Due to the suppression of weakly coupled vacua $s\gg1$ and $\mathcal{O}(1)$ values of the superpotential, the number of vacua that can be uplifted to dS via K\"ahler uplifting is strongly suppressed. For $\Delta A = 10$, only 6, i.e. a fraction of $\sim 10^{-4}$ of the total number of flux vacua allow such an uplifting.

The available tuning of the cosmological constant via fluxes can be estimated again via eq.~\eqref{cctuning}. The K\"ahler moduli stabilization yields a volume of $\Vol \simeq 50$~\cite{Louis:2012nb} such that the untuned cosmological constant is $\Lambda \sim 5 \cdot 10^{-6}$ and
\begin{equation}
 \frac{\Delta \Lambda}{\Lambda} \simeq 0.51 \pm 0.37\,,\label{DeltaLambdaKupliftparam}
\end{equation}
for $\Delta A = 10$. We remind the reader, that this is calculated for $L=34$ which is the maximal value we reach in our paramotopy scan and hence less than $L=104$ which is maximally allowed by the gauge flux and D7 brane construction realized for a K\"ahler uplifted dS vacuum in our explicit $\Pex$ example.

To summarize this section, the polynomial homotopy continuation method allows us to find all flux vacua for a given D3 tadpole $L$. The number of these vacua is well estimated by the statistical analysis of~\cite{Ashok:2003gk,Denef:2004ze}. We find that strongly coupled vacua $s \gtrsim 1$ are preferred as well as $\mathcal{O}(10^1-10^3)$ values of $W_0$. Our results can be used to estimate the tunability of the cosmological constant by fluxes and the number of flux vacua that can be K\"ahler uplifted to a dS vacuum.

\section{The minimal flux method} \label{minflux_sec}

In this section we describe the method to find flux vacua that has been first used by Denef, Douglas and Florea~\cite{Denef:2004dm}. In contrast to the polynomial homotopy continuation method described in section~\ref{paramotopy_sec}, we fix starting values for the VEVs $\langle U_1\rangle_{fix}$, $\langle U_2\rangle_{fix}$ and $\langle \tau\rangle_{fix}$ and solve for the flux values $f$ and $h$.

\subsection{The algorithm} \label{algo_minflux}

Due to the linear dependence of $W_0$ on $f$ and $h$, see eq.~\eqref{W0period}, the quantities $D_I W_0 = {W_0}_I + K_I W_0$ for $I=\tau,U_1,U_2$ are linear in these flux vectors. Hence, if we want to solve the system of equations
\begin{equation}
(W_0,D_{\tau}W_0,D_{U_1}W_0,D_{U_2}W_0)=0\,,\label{W8eq}
\end{equation}
this can be written as
\begin{equation}
 M\cdot (f,h) = 0\,,\label{Afh0}
\end{equation}
with $M \in \mathbb{R}^{8\times12}$ for general VEVs $\langle U_1\rangle, \langle U_2\rangle, \langle \tau\rangle \in \mathbb{C}$. The dimensions of $M$ are due to the fact that we have 8 real equations in eq.~\eqref{W8eq}, and there are 12 flux integers in total in $f$ and $h$. Note that we have also included the condition $W_0=0$ in eq.~\eqref{W8eq}. In fact, we are not interested in flux vacua where $W_0$ is strictly zero since none of the well studied moduli stabilization mechanisms KKLT~\cite{Kachru:2003aw}, LVS~\cite{Balasubramanian:2005zx} and K\"ahler uplifting~\cite{Balasubramanian:2004uy,Westphal:2006tn,Rummel:2011cd,Louis:2012nb} apply in this situation. However, eq.~\eqref{W8eq} will only serve as a starting point and we will eventually end up with vacua where $W_0 \neq 0$ and $\mathcal{O}(1)$.

For $M \in \mathbb{R}^{8\times12}$ there is no hope to find a solution of eq.~\eqref{Afh0} since the flux parameters have to be integers. However, if we neglect instanton corrections induced by eq.~\eqref{Ginst} and choose rational starting values $\langle U_1\rangle, \langle U_2\rangle, \langle \tau\rangle \in \mathbb{Q}+i\,\mathbb{Q}$ in the superpotential, the only transcendental number in eq.~\eqref{Afh0} is $\xi=\frac{\zeta(3)\chi}{2(2\pi\,i)^3}= -1.30843..\,i$. If we approximate $\xi$ by a rational number $\xi_{rat}$, for instance $\xi_{rat} = -13/10\,i$, we have accomplished $M \in \mathbb{Q}^{8\times12}$.

Now, we can hope to find a solution of eq.~\eqref{Afh0} although the entries of $f$ and $h$ will be generically be quite large for generic $M$, since one generally expects them to be at least of the order of the lowest common denominator of the entries of $M$. This puts tension on the D3 tadpole constraint eq.~\eqref{D3tad} since generally the geometry of the compactification manifold and D7 brane configuration generates $L \sim 10^2-10^4$. We use the same algorithms~\cite{Cohenbook1993} as the authors of~\cite{Denef:2004dm}, to generate as small as possible values for the entries of $f$ and $h$ in order to generate a not too large D3 tadpole $L < L_{max}$ where we choose $L_{max}=500$ to be the maximal value for the D3 tadpole that we consider.

Since the system of equations eq.~\eqref{Afh0} is under determined, the solution space is given by all linear combinations of linearly independent vectors $(f,h)_{i}$ for $i=1,..,4$, where each $(f,h)_{i}$ is a solution to eq.~\eqref{Afh0}, i.e.
\begin{equation}
 (f,h)_{sol}= \sum_{i=1}^4 a_i\, (f,h)_{i} \qquad \quad \text{with } a_i \in \mathbb{Z}\,.
\end{equation}
For obvious practical reasons, we cannot consider all possible values of the $a_i$. Since $L(a_i (f,h)_{i}) = a_i^2 L((f,h)_{i})$ and tadpoles of solution basis vectors smaller than ten are extremely rare, we can safely restrict ourselves to $-3<a_i<3$ in order to fulfill $L((f,h)_{sol})<L_{max}$.

In the next step, we are looking for solutions to eq.~\eqref{TOSOLVE}, including instanton corrections eq.~\eqref{Ginst} to the prepotential and also setting $\xi$ to its transcendental value. We insert the flux solution $(f,h)_{sol}$ into the equations
\begin{equation}
 (D_{\tau}W_0,D_{U_1}W_0,D_{U_2}W_0)=0\qquad\quad \text{(with instanton corrections)}\,,
\end{equation}
leaving the VEVs $\langle U_1\rangle$, $\langle U_2\rangle$ and $\langle \tau\rangle$ unfixed. These are six real equations for six real variables and we can numerically search for a solution in the vicinity of $\langle U_1\rangle_{fix}$, $\langle U_2\rangle_{fix}$ and $\langle \tau\rangle_{fix}$. It has to be checked case by case if the complex structure limit is still valid for these perturbed solutions. The shift of the VEVs from their fixed values may also induce a shift in the superpotential, i.e. $W_0$ is not zero anymore. However, note that the value that $W_0$ will take in the end is not in any way under our control and it has to be checked if one obtains useful values for the purpose of moduli stabilization.

The above outlined algorithm can be iterated by sampling over a set of VEVs $\langle U_1\rangle_{fix}$, $\langle U_2\rangle_{fix}$ and $\langle \tau\rangle_{fix}$ and approximate $\xi$ values $\xi_{rat}$.

\subsection{The scan}

First of all, we need to define a set of rational starting values $(\langle U_1\rangle,\langle U_2\rangle,\langle \tau\rangle,\xi_{rat})_{fix}$ over which the algorithm explained in section~\ref{algo_minflux} can be iterated. We set the axionic components of the fixed VEV's to zero, i.e.
\begin{equation}
\text{Re}\langle U_1\rangle_{fix} = \text{Re}\langle U_2\rangle_{fix} = \text{Re}\langle \tau\rangle_{fix} =0\,.
\end{equation}
Let $x\in[x_{min},x_{max}]$ represent $\xi_{rat}$ and the imaginary parts of the moduli. We use
\begin{equation}
 \left\{\frac{p}{q}\,\,| \,\, p\in[\text{min}_{\mathbb{Z}}(q\,x_{min}),\text{max}_{\mathbb{Z}}(q\,x_{max})]\,,q\in[1,q_{max}] \right\}\,,\label{pqformula}
\end{equation}
as the set of rational numbers that fills out the interval $[x_{min},x_{max}]$.  $\text{min}_{\mathbb{Z}}(x)$ and $\text{max}_{\mathbb{Z}}(x)$ define the smallest integer greater than or equal to $x$ respectively the greatest integer less than or equal to $x$. Note that $q_{max}$ determines how `dense' the interval is filled with rational numbers. For example, one has
\begin{equation}
 \left\{1,\frac{5}{4},\frac{4}{3},\frac{3}{2},\frac{5}{3},\frac{7}{4},2\right\}
\end{equation}
for $x \in [1,2]$ and $q_{max}=4$.

\begin{table}[ht!]
\centering
  \begin{tabular}{|c|c|c|c|c|}
  \hline
   & $x_{min}$ & $x_{max}$ & $q_{max}$ & $\#$\\
  \hline
  $\xi_{rat}$ & $0.9\xi$ & $1.1\xi$ & $20$ & $34$\\
  \hline
  $u_1$ & $1$ & $4$ & $5$ & $31$\\
  \hline
  $u_2$ & $1/6$ & $4$ & $5$ & $40$\\
  \hline
  $s$ & $1$ & $10$ & $10$ & $289$\\
  \hline
  \end{tabular}
  \caption{Starting values for $\xi_{rat}$ and the imaginary parts of the moduli. The last column counts the number of elements yielding from the choice of $x_{min}$, $x_{max}$ and $q_{max}$ according to eq.~\eqref{pqformula}.}
  \label{startingvalues_tab}
\end{table}
The set of starting values given in Table~\ref{startingvalues_tab} is chosen such that the starting values for the complex structure moduli are in the large complex structure limit eq.~\eqref{largecompstr} and the string coupling $g_s = 1/s$ is always larger than one. The total number of points in the grid of starting values defined in Table~\ref{startingvalues_tab} has 12,184,240 points. Using 80 cores with 2.4 GHz of the \textit{DESY Theory Cluster} this yields a total calculation time of approximately four weeks.

\subsection{Results}

\begin{minipage}{0.49\linewidth}
We find 1,698 solutions fulfilling the constraint $L<L_{max}=500$ which is only $0.01\%$ of the total number of points of the scan. Hence, most of the time one can not find a flux vector whose entries are small enough to fulfill the D3 tadpole constraint.

As in section~\ref{distr_param_sec}, we can make use of the $SL(2,\mathbb{Z})$ transformations eq.~\eqref{S+btrans} and eq.~\eqref{-1Strans} to transform every solution to the fundamental domain eq.~\eqref{funddomain}. Identifying equivalent solutions in this domain, 1,374 elements of the original solution set are not related via $SL(2,\mathbb{Z})$ symmetry and hence physically inequivalent. The distribution of $\tau$ is shown in Figure~\ref{dilatondistr_fig}. Compared to the paramotopy scan we more easily find weakly coupled vacua with $s\gg1$ which is due to the fact that we have chosen the starting values for $s$ accordingly, see Table~\ref{startingvalues_tab}.
\end{minipage}
\hskip 5mm
\begin{minipage}{0.45\linewidth}
\begin{center}
\includegraphics[width= \linewidth]{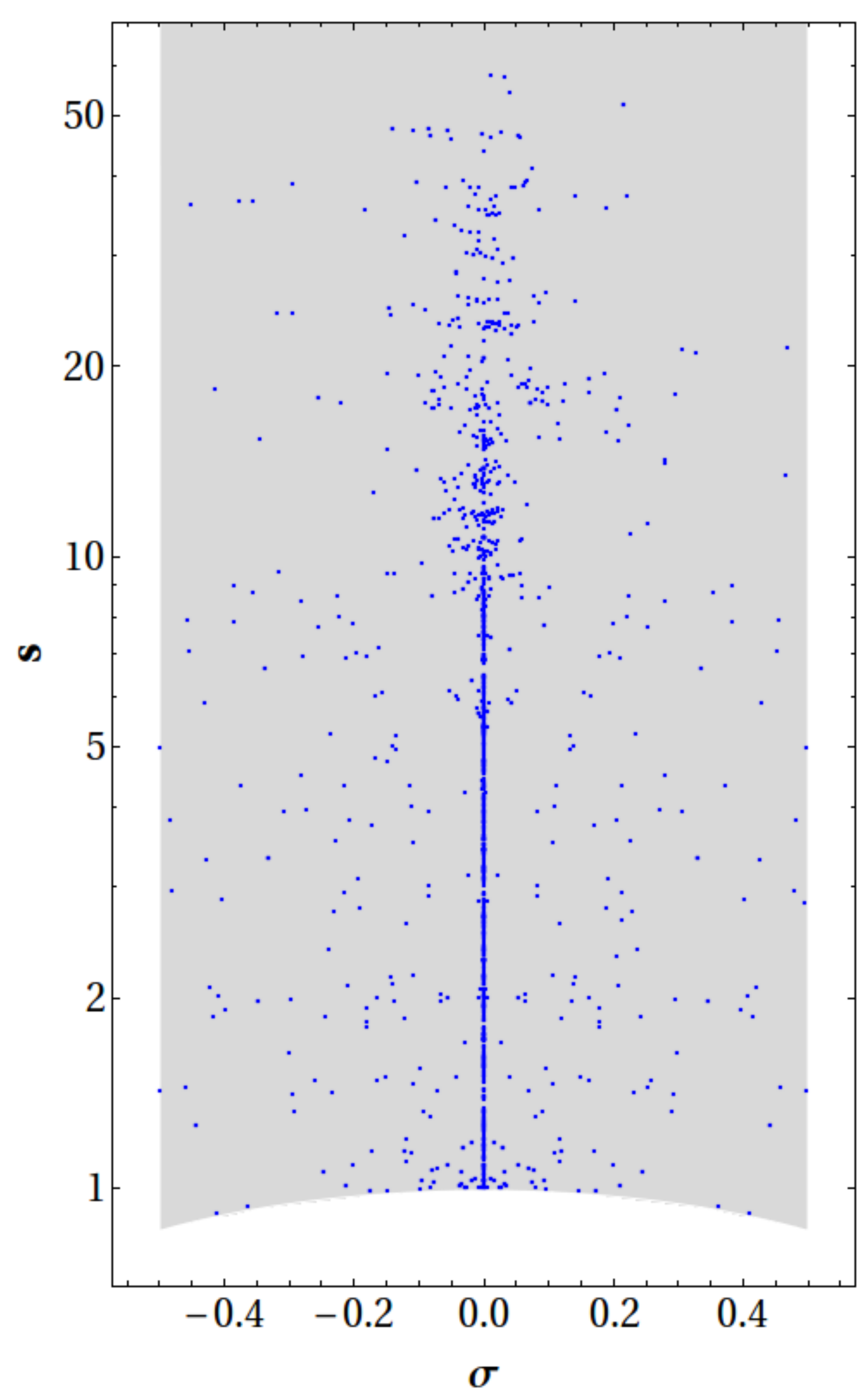}
\captionof{figure}{Distribution of $\tau$ for the minimal flux scan.}
\label{dilatondistr_fig}
\end{center}
\end{minipage}
\vskip 2mm

\begin{figure}[t!]
\centering
\includegraphics[width= 0.48\linewidth]{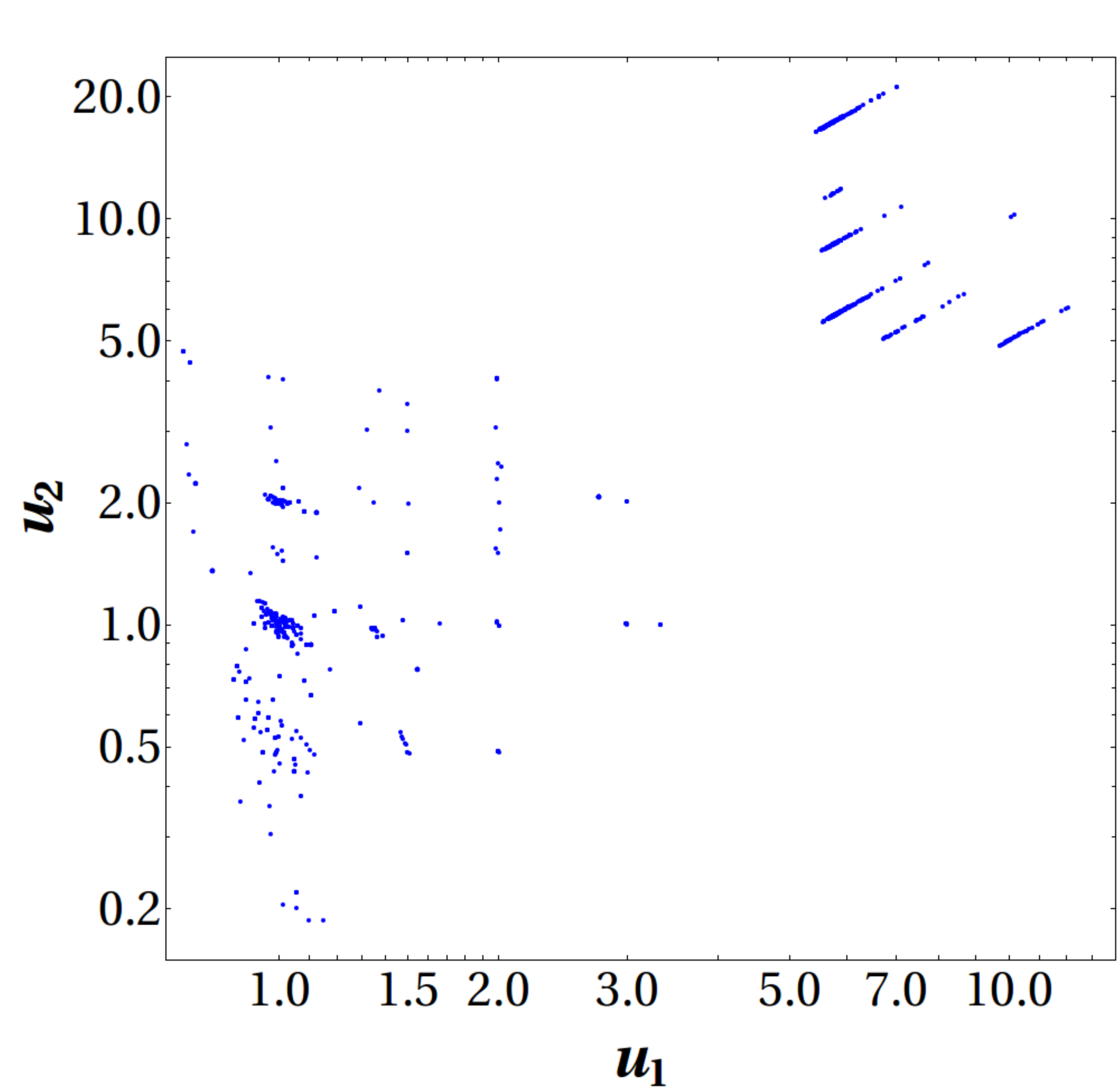}
\hskip 3mm
\includegraphics[width= 0.48\linewidth]{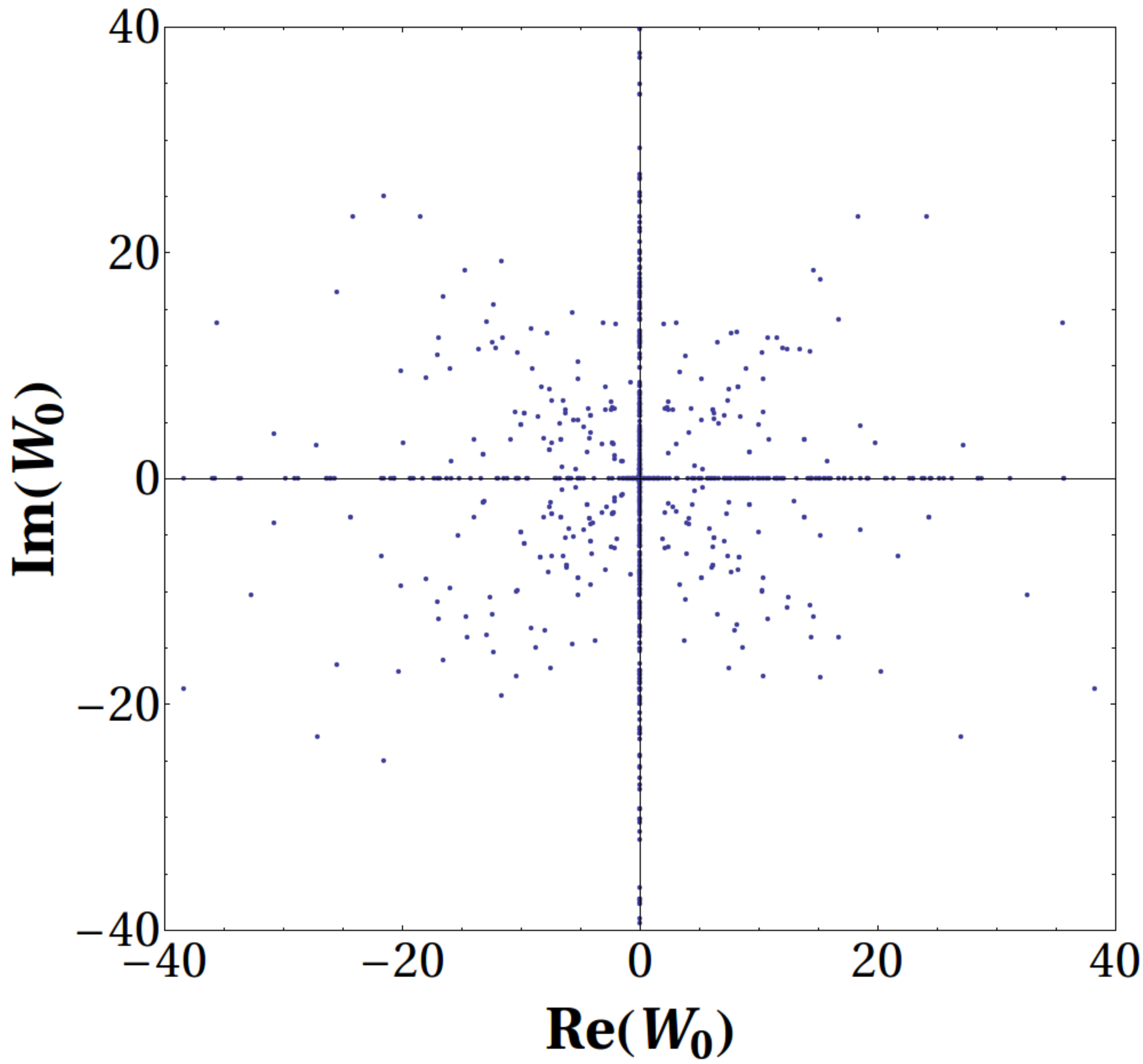}
\caption{Distribution of $u_1$ and $u_2$ (left) and superpotential $W_0$ for the minimal flux scan.}
\label{LCSW0minflux_fig}
\end{figure}

The distribution of $u_1$ and $u_2$ as well as the distribution of the superpotential are shown in Figure~\ref{LCSW0minflux_fig}. We find that there are no flux vacua in the vicinity of the conifold singularities eq.~\eqref{CFcriterium} for $\epsilon_{CF} = 10^{-2}$. Also, all flux vacua fulfill the constraint of the validity of the large complex structure limit description, eq.~\eqref{largecompstr} for $\epsilon_{LCS} = 10^{-2}$ which is again due to the chosen starting values deep in the large complex structure limit, Table~\ref{startingvalues_tab}. Since we solve for vanishing $W_0$ in the first step of the algorithm eq.~\eqref{W8eq}, we obtain a clustering around $W_0 = 0$, see Figure~\ref{LCSW0minflux_fig}. This is however not a general property of the complete solution space as we noted in section~\ref{distr_param_sec} but rather $\mathcal{O}(10-100)$ values are preferred.

We fit the number of vacua as a function of the D3 tadpole $L$, finding $N_{vac} \simeq 0.02\, L^{1.83}$ which strongly deviates from the estimate eq.~\eqref{NvacDenefDouglas}. However, our dataset of 1,374 is in no way representative for the total number of flux vacua with $L=500$ such that this deviation can be easily explained by insufficient statistics.

\begin{figure}[t!]
\centering
\includegraphics[width= 0.49\linewidth]{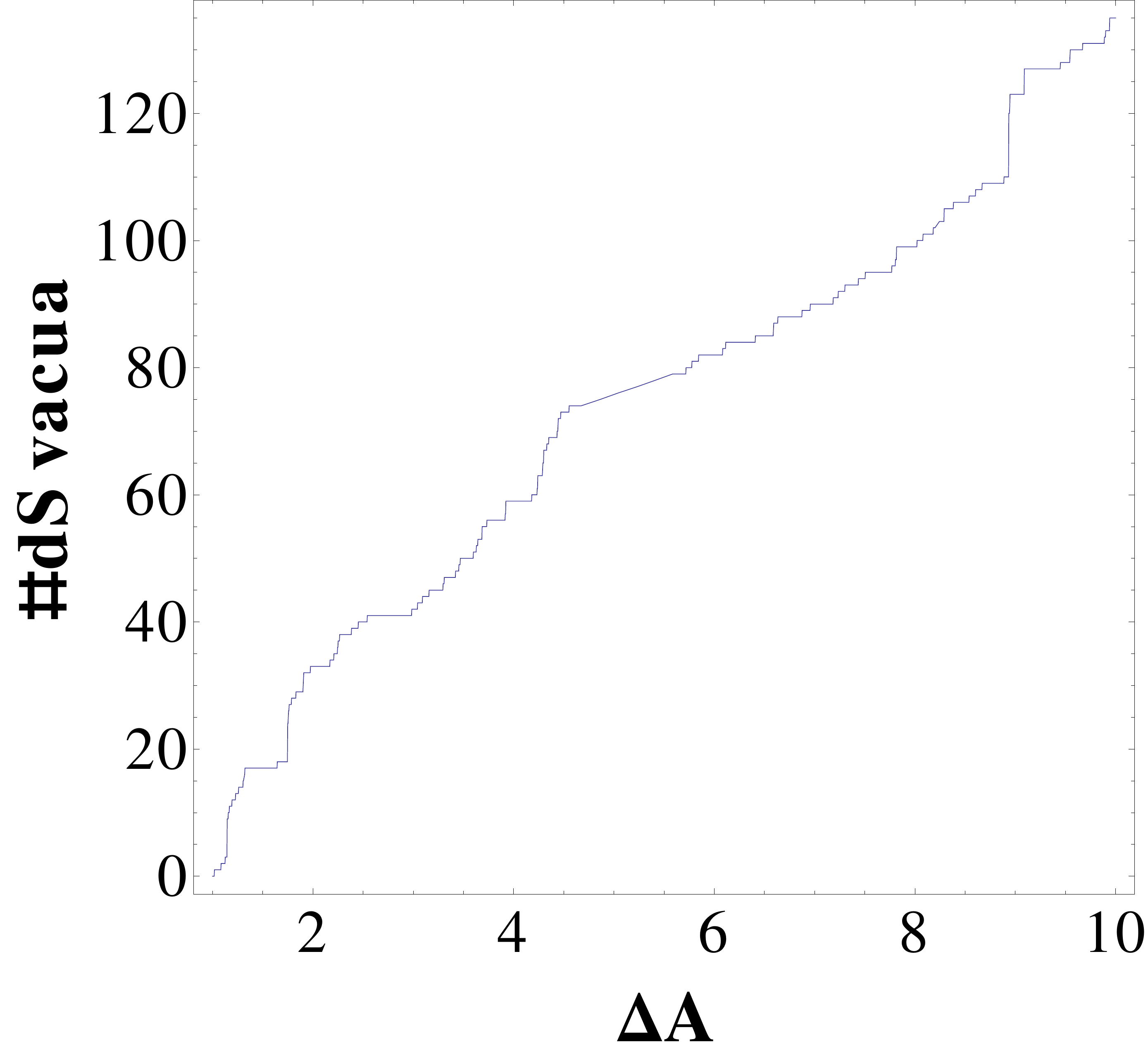}
\includegraphics[width= 0.49\linewidth]{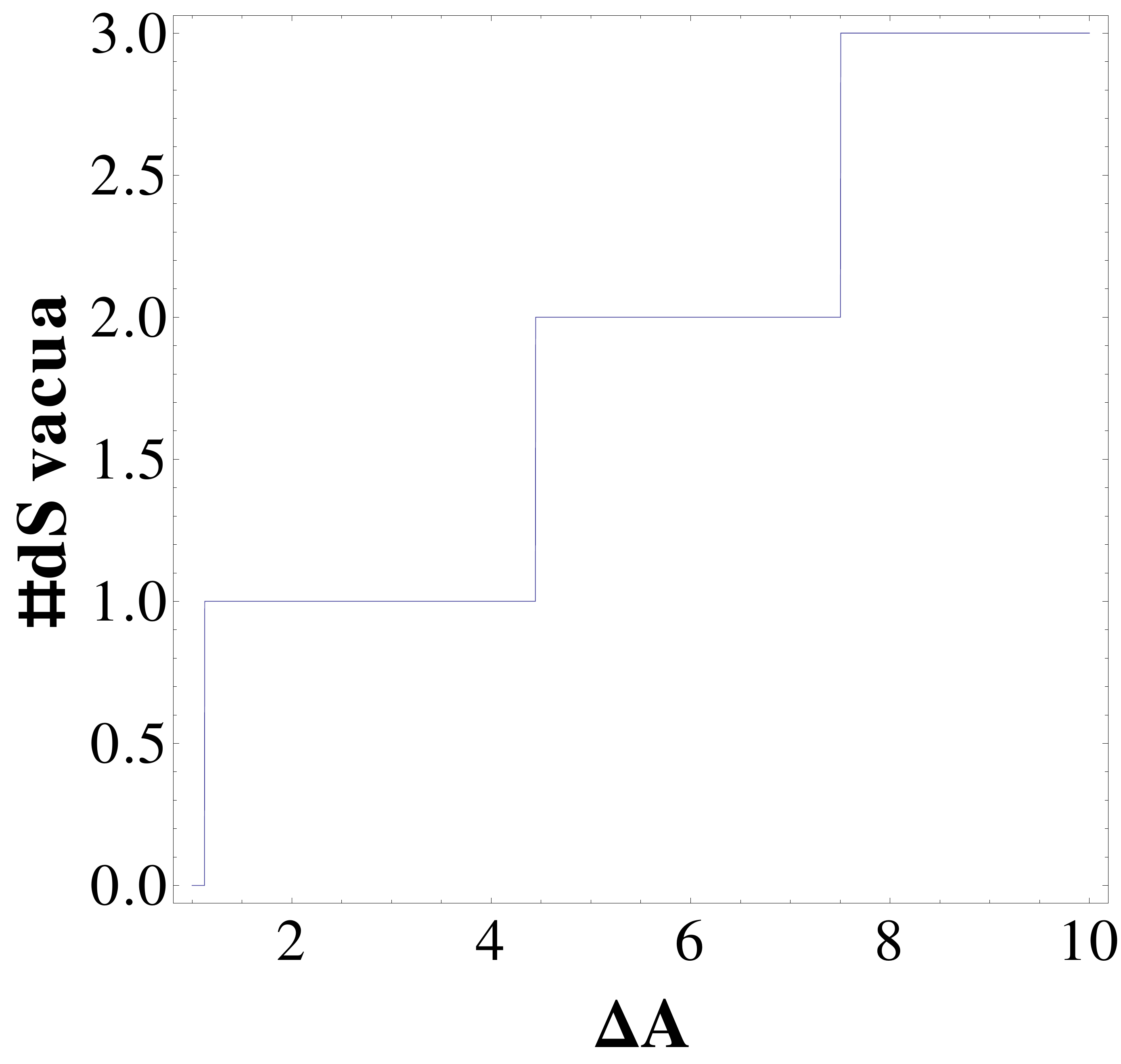}\\[2mm]
\includegraphics[width= 0.49\linewidth]{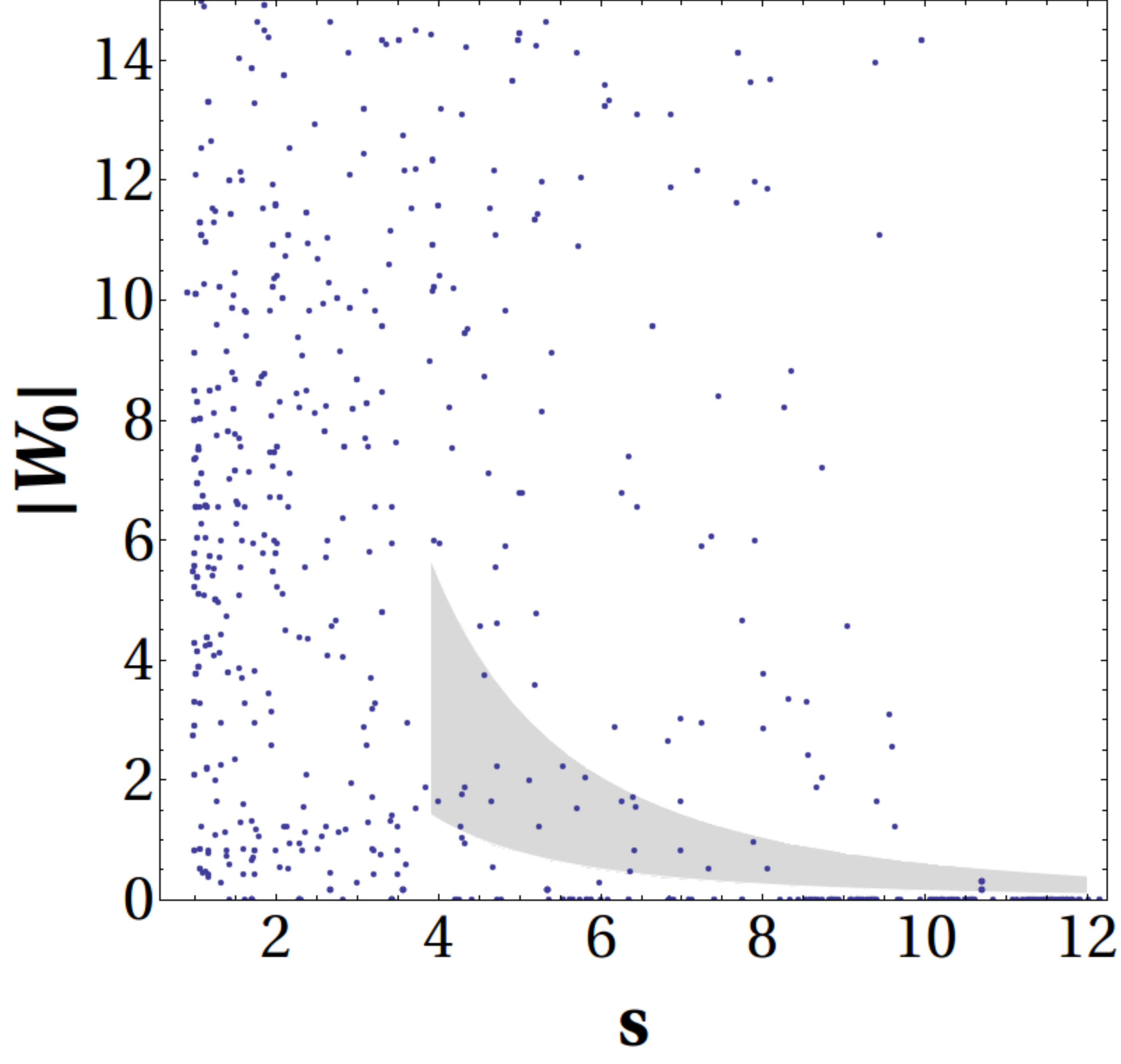}
\includegraphics[width= 0.49\linewidth]{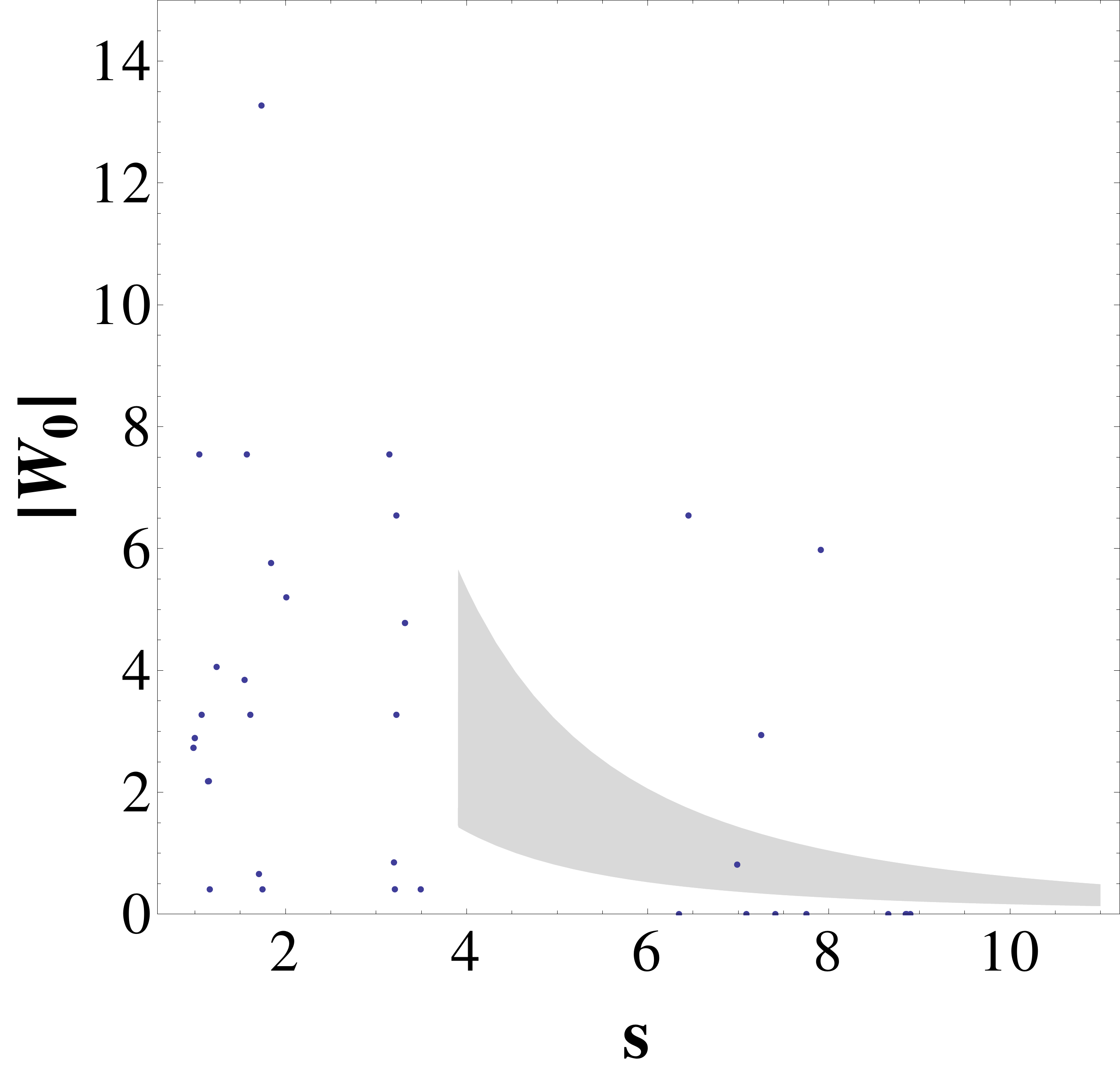}
\caption{The number of K\"ahler uplifted dS vacua as a function of $\Delta A$ (top) and data points $(s,|W_0|)$ (bottom). K\"ahler uplifting can be applied in the shaded region ($\Delta A =2$) for the minimal flux scan. We show the results for $L=500$ (left) and $L=104$ (right).}
\label{dScountregion_fig}
\end{figure}

Finally, we can calculate how many flux vacua allow a dS vacuum via K\"ahler uplifting along the lines of section~\ref{deSitter_param_sec}. Since the minimal flux scan is setup such that the values of $s$ and $|W_0|$ naturally lie in the region where K\"ahler uplifting can be applied we find a much milder suppression of these vacua compared to section~\ref{distr_param_sec}, see Figure~\ref{dScountregion_fig}. For $\Delta A = 10$, three of the 75 flux vacua with $L=104$ and 135 of the 1,374 flux vacua with $L=500$ allow a dS vacuum via K\"ahler uplifting.\footnote{The maximum D3 tadpole of $X_3$ is $L=104$ but due to the small amount of vacua we find for this tadpole we also show the results for $L=500$.} Repeating the estimate of eq.~\eqref{DeltaLambdaKupliftparam} for the results of the minimal flux scan, the cosmological constant $\Lambda \sim 10^{-6}$ can be tuned to an accuracy
\begin{eqnarray}
\begin{aligned}
&\frac{\Delta \Lambda}{\Lambda} \sim 1.9\pm 2.2\,\qquad &\text{for} \qquad L=104\,,\\
&\frac{\Delta \Lambda}{\Lambda} \sim 5\cdot 10^{-2}\pm 10^{-2}\,\qquad &\text{for} \qquad L=500\,.\\
\end{aligned}
\end{eqnarray}

To conclude this section, the minimal flux method has the advantage that one can specify the region in parameter space where physically interesting flux vacua should be constructed. In our case this region is defined by the large complex structure limit, weak string coupling and $\mathcal{O}(1)$ values of $W_0$. However, it is not possible to construct all flux vacua for a given D3 tadpole $L$ which can be done using paramotopy and the method is rather inefficient in the sense that only $0.01\%$ of the starting values yield a flux vacuum.

\section{Conclusions} \label{conclusions}

In this paper we have studied the flux vacua of type IIB string theory compactified on $\Pex$ Calabi-Yau hypersurface, i.e. the standard working example of both the LVS and the K\"ahler uplifting scenario. We switch on flux along six three-cycles that correspond to two complex structure moduli that are invariant under a certain discrete symmetry that can be used to construct the mirror manifold. As explained in the main text,  such a supersymmetric vacuum in these two complex structure moduli extends to a supersymmetric vacuum of \textit{all} 272 complex structure moduli.

To explicitly construct flux vacua, we make use of the fact that the prepotential $\mathcal{G}$ of the two complex structure moduli space has been worked out in the large complex structure limit. We apply two computational methods to find flux vacua on this manifold: The polynomial homotopy continuation method allows us to explicitly construct for the first time \textit{all flux vacua} in the large complex structure limit that are consistent with a given D3 tadpole $L$ by applying the polynomial homotopy continuation method at each point in flux parameter space. The minimal flux method finds flux parameters that are consistent with a given D3 tadpole $L$ for a given set of vacuum expectation values (VEVs) of the complex structure moduli.

We analyze the resulting solution space of flux vacua for several physically interesting properties. For the polynomial homotopy continuation method, we find that for the $\sim 50,000$ parameter points of our scan there are $\sim 20,000$ solutions in the large complex structure limit. We find a preference of strongly coupled vacua $g_s \gtrsim 1$ and preference for values of $\mathcal{O}(10^1-10^3)$ for the flux superpotential $W_0$. The number of vacua is
\begin{equation}
N_{vac} \simeq (0.52\pm 0.04)\, L^{2.92\pm 0.03} \,,
\end{equation}
compared to $\sim 0.03 L^3$ expected form statistical analysis~\cite{Ashok:2003gk,Denef:2004ze}. The gravitino mass is typically $m_{3/2}^2=\mathcal{O}(10^{-3}) \cdot(\frac{100}{\Vol})^{2} M_{\text{P}}^2$ and the masses of the complex structure moduli and the dilaton scale like $\mathcal{O}(10^{-3}-10^2) \cdot(\frac{100}{\Vol})^{2} M_{\text{P}}^2$. These ranges for the moduli and gravitino masses are compatible with the values obtained for a single explicit flux choice in the same construction~\cite{Rummel:2011cd,Louis:2012nb}.

The average spacing of the flux superpotential in our solution set can be used to estimate the available fine-tuning $\Delta \Lambda / \Lambda$ of the cosmological constant $\Lambda$ as
\begin{equation}
 \frac{\Delta \Lambda}{\Lambda} \simeq (5.1 \pm 0.3)\,L^{- (0.93\pm 0.006)\, ( h^{2,1}_{\text{eff}}+1)}\,,
\end{equation}
which corresponds to for instance $\Delta \Lambda / \Lambda \sim 10^{-100}$ for $L=500$ and $h^{2,1}_{\text{eff}} = 40$. The explicit brane and gauge flux construction in~\cite{Louis:2012nb} allows us to answer the question how many of theses supersymmetric flux vacua allow an uplift to dS via K\"ahler uplifting. Depending on the available values for the one-loop determinant from gaugino condensation used to stabilize the K\"ahler moduli in this setup, we find that for a fraction of about $10^{-4}$ of all flux vacua up to a given D3-brane tadpole this mechanism can be applied to obtain a dS vacuum.

For the minimal flux method, we find $\sim 1000$ flux vacua with $L<500$ out of $\sim 10^7$ parameter points of our scan. This method allows us to control the region in $W_0$ and moduli space where we are intending to find flux vacua. Hence, we more easily access the regions of weak string coupling and the large complex structure limit compared to the polynomial homotopy continuation method. For the much smaller set of flux vacua constructed with the minimal flux method, the fraction of K\"ahler uplifted dS minima is about $10\%$.

\acknowledgments We thank Frederik Denef for the initial motivation in launching this project, and Jose Blanco-Pillado, Arthur Hebecker, Jan Louis, Francisco Pedro, and Roberto Valandro for valuable, and enlightening discussions. MR thanks Malte Nuhn for help with Python. This work was supported by the Impuls und Vernetzungsfond of the Helmholtz Association of German Research Centers under grant HZ-NG-603, the German Science Foundation (DFG) within the Collaborative Research Center 676 "Particles, Strings and the Early Universe" and
the Research Training Group 1670. DM would like to thank the U.S. Department of Energy
for their support under contract no. DE-FG02-85ER40237; Fermilab's LQCD clusters where the cheater's homotopy runs were performed; and most importantly, Matthew Niemerg for his continuous support in installing and help in running Paramotopy. Finally, the authors would like to thank the organizers of the 2012 String Phenomenology conference in Cambridge where part of the work was done for their warm hospitality.

\bibliographystyle{JHEP.bst}
\bibliography{fluxvacua}
\end{document}